\documentclass[journal]{IEEEtran}

\usepackage[dvipsnames]{xcolor}
\usepackage[acronym]{glossaries}
\usepackage[colorlinks, citecolor=blue, linkcolor=blue]{hyperref}
\usepackage[textwidth=1.6cm]{todonotes}
\usepackage{comment}
\usepackage{nameref}
\usepackage{subcaption}
\usepackage{multicol}
\usepackage{multirow}
\usepackage{rotating}
\usepackage{stfloats}
\usepackage[noadjust]{cite}
\usepackage{float}
\usepackage{amsfonts}
\usepackage{upgreek}
\usepackage{mathtools}
\usepackage{scalerel}
\usepackage{braket}
\usepackage{soul}
\usepackage{tcolorbox}
\DeclareMathOperator{\Tr}{Tr}
\newcommand{\Th}{\mathrm{th}}
\newcommand{\Int}{\mathrm{Int}}
\newcommand{\Sig}{\mathrm{Sig}}
\newcommand{\Tx}{\mathrm{Tx}}
\newcommand{\Total}{\mathrm{T}}
\newcommand{\Real}[1]{\mathbb{R}\left\{ #1\right\}}
\newcommand{\R}{\mathrm{R}}

\newcommand{\Max}{\mathrm{Max}}
\newcommand{\Min}{\mathrm{Min}}

\newcommand{\Fwd}{\mathrm{F}}
\newcommand{\Bwd}{\mathrm{B}}

\newcommand{\Leff}{L_{\mathrm{eff}}}
\newcommand{\Aeff}[1]{A_{\mathrm{eff#1}}}
\newcommand{\Evec}{\vec{\mathbf{E}}}
\newcommand{\Avec}{\vec{\mathbf{A}}}
\newcommand{\Nch}{N_{\mathrm{ch}}}
\newcommand{\der}{\mathrm{d}}

\let\bigopsize\bigoplus
\def\bigoplus{{\scalerel*{\boldsymbol\oplus}{\bigopsize}}}

\newcommand{\appref}[1]{\hyperref[#1]{appendix~\ref*{#1}}}
\newcounter{MYtempeqncnt}
\makeatletter
\newcommand{\subalign}[1]{%
  \vcenter{%
    \Let@ \restore@math@cr \default@tag
    \baselineskip\fontdimen10 \scriptfont\tw@
    \advance\baselineskip\fontdimen12 \scriptfont\tw@
    \lineskip\thr@@\fontdimen8 \scriptfont\thr@@
    \lineskiplimit\lineskip
    \ialign{\hfil$\m@th\scriptstyle##$&$\m@th\scriptstyle{}##$\hfil\crcr
      #1\crcr
    }%
  }%
}
\makeatother
\newacronym{snr}{SNR}{signal-to-noise ratio}
\newacronym{sinr}{SINR}{signal-to-interference-plus-noise ratio}
\newacronym{sdm}{SDM}{space-division multiplexing}
\newacronym{mdg}{MDG}{mode-dependent gain}
\newacronym{mdl}{MDL}{mode-dependent loss}
\newacronym{csi}{CSI}{channel state information}
\newacronym{cdf}{CDF}{cumulative distribution function}
\newacronym{pdf}{PDF}{probability density function}
\newacronym{gue}{GUE}{Gaussian unitary ensemble}
\newacronym{mmse}{MMSE}{minimum mean squared error}
\newacronym{mimo}{MIMO}{multiple-input multiple-output}
\newacronym{msle}{MSLE}{mean squared logarithmic error}
\newacronym{awgn}{AWGN}{additive white Gaussian noise}
\newacronym{nlse}{NLSE}{non-linear Schrödinger equations}
\newacronym{mmf}{MMF}{multi-mode fiber}
\newacronym{mcf}{MCF}{multi-core fiber}
\newacronym{ssmf}{SSMF}{standard single-mode fiber}
\newacronym{psd}{PSD}{power spectral density}
\newacronym{rk4}{RK4}{Runge-Kutta 4$^{\Th}$ order}
\newacronym{snu}{SNU}{shot noise units}

\newacronym{lp}{LP}{linearly-polarized}
\newacronym{qkd}{QKD}{quantum-key distribution}
\newacronym{skr}{SKR}{secret key rate}
\newacronym{ase}{ASE}{amplified spontaneous emission}
\newacronym{srs}{SRS}{stimulated Raman scattering}
\newacronym{sprs}{SpRS}{spontaneous Raman scattering}
\newacronym{spm}{SPM}{self-phase modulation}
\newacronym{xpm}{XPM}{cross-phase modulation}
\newacronym{fwm}{FWM}{four-wave-mixing}
\newacronym{imxt}{IMXT}{inter-mode crosstalk}
\newacronym{qber}{QBER}{quantum bit error rate}

\begin{document}
\title{Accurate and Effective Model for Coexistence of Classical and Quantum Signals In Optical Fibers}
\author{Lucas~Alves~Zischler,~\IEEEmembership{Student~Member,~IEEE,}%
        ~Çağla~Özkan,~\IEEEmembership{Student~Member,~IEEE,}
        ~Tristan~Vosshenrich,~\IEEEmembership{Student~Member,~IEEE,}%
        ~Qi~Wu,~\IEEEmembership{Student~Member,~IEEE,}
        ~Giammarco~Di~Sciullo,~\IEEEmembership{Student~Member,~IEEE,}%
        ~Divya~A.~Shaji,~\IEEEmembership{Student~Member,~IEEE,}
        ~Chiara~Lasagni,~\IEEEmembership{Member,~IEEE,}%
        ~Paolo~Serena,~\IEEEmembership{Senior~Member,~IEEE,}
        ~Alberto~Bononi,~\IEEEmembership{Senior~Member,~IEEE,}%
        ~Amirhossein~Ghazisaeidi,~\IEEEmembership{Senior~Member,~IEEE,}
        ~Chigo~Okonkwo,~\IEEEmembership{Senior~Member,~IEEE,}%
        ~Antonio~Mecozzi,~\IEEEmembership{Fellow,~IEEE,~Optica,}
        ~and~Cristian~Antonelli,~\IEEEmembership{Senior~Member,~IEEE,~Fellow,~Optica}
% Lower left notes
\thanks{Manuscript received XXX xx, XXXX; revised XXXXX xx, XXXX; accepted XXXX XX, XXXX. This work was supported in part by the European Union's Grant Agreement No. 101120422 - Quantum Enhanced Optical Communication Network Security (QuNEST) and No. 101072409 - Optical Fiber Higher Order mode Technologies (HOMTech). (Corresponding Author: Lucas~Alves~Zischler)}%
\thanks{Lucas Alves Zischler, Giammarco Di Sciullo, Divya A. Shaji, ${\text{Antonio Mecozzi}}$, and Cristian Antonelli are with the Department of Physical and Chemical Sciences, University of L’Aquila, 67100 L’Aquila, Italy: (\mbox{e-mail: lucas.zischler@univaq.it}).}%
\thanks{Çağla Özkan and Chigo Okonkwo are with the High Capacity Optical Transmission Laboratory, Electro-Optical Communications Group, Eindhoven University of Technology, 5600 MB Eindhoven, The Netherland.}%
\thanks{Tristan Vosshenrich and Amirhossein Ghazisaeidi are with the Nokia Bell Labs, 91300 Massy, France.}
\thanks{Chiara Lasagni, Paolo Serena, and Alberto Bononi are with with the Department of Engineering and Architecture, Università Degli Studi di Parma, 43124 Parma, Italy.}%
\thanks{Qi Wu is with the State Key Laboratory of Advanced Optical Communication Systems and Networks, Department of Electronic Engineering, Shanghai Jiao Tong University, Shanghai, 200240, China.}}%

% Paper headers
\markboth{}%
{Accurate and Effective Model for Coexistence of Classical and Quantum Signals In Optical Fibers}

% Title area
\maketitle

% Abstract
\begin{abstract}
  The rising interest in quantum-level communication has resulted in proposals for coexistence schemes with classical signals within the same fiber optic channel, where the most recent proposals leverage novel fibers designed for \gls*{sdm} transmission. In all cases the large power difference between classical and quantum channels presents challenges for such schemes, as the classical signals generate interfering noise that corrupts the quantum signal. In this work, we discuss  the main interference mechanisms in coexistence scenarios and provide a model to quantify their impact on the quantum signal quality. Analytical approximations in the model allow accurate and fast numerical solutions in the millisecond time-scale. The model accounts for out-of-band non-linear interference effects, namely \gls*{sprs} and \gls*{fwm} in both cases of single-mode and \gls*{sdm} fibers with weakly-coupled degenerate mode groups. Rayleigh and \gls*{sprs} backscattering are considered in counter-propagating scenarios. Since broadband classical transmission is targeted, the model also accounts for the effect of \gls*{srs}-induced power tilt. Use of the model in sample scenarios indicates that the interference noise power is minimized at the high end of the transmission band in both cases were the quantum is co- and counter-propagating with respect to the classical signals, with a preference of one or the other scheme depending on the link length and quantum signal center frequency. Our model reveals that \gls*{fwm} has negligible impact in counter-propagating schemes, but can be relevant in co-propagating schemes under certain scenarios. Nevertheless, the \gls*{fwm} interference can be mitigated by deallocating the classical signals adjacent to the quantum channel.
\end{abstract}

% Keywords

% Glossary
\glsresetall

% Section: Introduction
\section{Introduction}

\IEEEPARstart{N}{ovel} communication protocols and applications are currently being proposed to harness the properties of quantum mechanics. Some works have proposed the transmission of information via quantum states of individual photons for communication, metrology applications, large-scale quantum computing, or cryptography~\cite{kimble2008quantum,lo2014secure}. Communication at the quantum level has become particularly relevant in cryptography, with \gls*{qkd} protocols~\cite{bennet1984quantum}. Current cryptographic protocols operate at the application layer and rely on the complexity of specific mathematical operations for security. However, the physical layer is left unprotected and is often assumed to be vulnerable to eavesdroppers. While current devices are not capable of breaking current cryptography within a viable timespan, algorithms designed for quantum computers are known to break some encryption protocols with ease~\cite{mavroeidis2018impact}.

Although no currently known quantum computer has sufficient qubits to break current encryption schemes, data being transmitted today can be stored until such a device is developed, which is an attack within the feasible range for state-level actors. To protect the cipher against such attacks, many post-quantum cryptographic protocols have been proposed. However, there are still questions raised about their security~\cite{beullens2022breaking,alagic2022status}. Another discussed approach is to provide security against eavesdroppers at the physical layer. This is the premise of \gls*{qkd} schemes, which rely on fundamental information limits of quantum physics to ensure that two or more trusted parties communicate without any leakage of information to an untrusted party~\cite{renner2005security}. Nevertheless, to leverage such quantum properties, the signal must be transmitted at quantum power levels.

Many quantum-level applications encode information in the states of photons, enabling the transmission of quantum signals through optical fibers. As a far-reaching optical network for classical data transmission is already in place, many works have explored the coexistence of classical and quantum signals on deployed links. Such coexistence schemes avoid the additional expenses of installing dedicated infrastructure for quantum transmission. Nevertheless, the power levels used in quantum and classical communications differ by orders of magnitude, and interference considered negligible in classical systems can result in significant impairments to quantum signals~\cite{diamanti2016practical}.

Quantum signal degradation can arise from additive interference at the transceivers, as well as from non-linear interference and spatial crosstalk of classical signals into quantum channels during transmission. The interference at the transceivers results from insufficient notch filtering of classical signals and \gls*{ase} noise while multiplexing the quantum signal in the frequency grid, in addition to additive noise and distortions introduced by the quantum devices. Interference during propagation are caused by physical interactions within the optical waveguide. In this study, we do not address interference occurring at the transmitter or receiver, as it is influenced by factors such as bandpass filter isolation and thermal noise~\cite{patel2012coexistence}. Instead, we concentrate solely on interference effects that occur during propagation. Despite the high sensitivity of quantum signals, numerous studies have demonstrated that their coexistence with classical signals is possible in real-world scenarios. As an example, in~\cite{townsend1997simultaneous,nakazawa2017qam,huang2015continuous,eriksson2018joint,eriksson2018coexistence,milovanvcev2020spectrally}, quantum and classical signals where transmitted simultaneously through~\gls*{ssmf}.

\Gls*{sdm}, which in recent years has been at the spotlight of optical communications as one of the most effective approach for scaling the capacity of future fiber-optic systems, enables new coexistence schemes by introducing an additional degree of freedom in the spatial dimension. Quantum signals can be transmitted through dedicated weakly coupled mode groups in \glspl*{mmf} or dedicated cores in \glspl*{mcf}, providing an extra layer of isolation. However, coexistence interference in \gls*{sdm} fibers require dedicated analysis due to the effects of spatial crosstalk. Several works have already investigated coexistence in \gls*{sdm} systems. For example,~\cite{eriksson2019inter,hugues202011,geng2022integration,dou2023coexistence,kong2024resource} present experimental evaluations of coexistence within \gls*{sdm} fibers.

Several works have also provided analytical models to evaluate the feasibility of coexistence. However, no existing model fully incorporates all interference contributions analytically, but rather discuss effects qualitatively or offer models tailored to specific scenarios. In~\cite{peters2009dense}, it is discussed that in \gls*{ssmf} the main non-linear interference contributions are from \gls*{sprs} and \gls*{fwm}, with some analytical insights provided. However, the authors' goal is to evaluate non-linear effects rather than to present a comprehensive model. In~\cite{eraerds2010quantum} and~\cite{kawahara2011effect}, the authors provide analytical models that accounts for \gls*{sprs} interference, supported by experimental measurements of the Raman cross-section profile, but \gls*{fwm} is not considered. In~\cite{niu2018optimized}, an analytical discussion of the combined effects of \gls*{sprs} and \gls*{fwm} is presented, and an optimized quantum channel placement is proposed to reduce non-linear interference. It is shown that an interleaved channel placement reduces \gls*{fwm} interference. However, the work is limited to unmodulated continuous waves. In~\cite{du2020impact}, analytical expressions for \gls*{sprs} and \gls*{fwm} are provided, but they are derived for a single-mode scenario with Gaussian-shaped signals and a flat attenuation profile.

Some works also analytically evaluate coexistence in \gls*{sdm} links. In~\cite{cai2020intercore}, analytical models accounting for the combined interactions of \gls*{sprs} and spatial crosstalk are provided. However, the frequency dependence of the attenuation coefficients are not correctly accounted, and \gls*{fwm} is disregarded. In~\cite{wu2025integration}, we propose an alytical models that include \gls*{sprs} and spatial crosstalk in a \gls*{mcf} setup, where the respective coefficient profiles are experimentally measured, but does not consider \gls*{fwm} interference neither backscattering Rayleigh in counter-propagating scenarios.

In this work, we present a comprehensive analysis of the key physical phenomena in optical fibers that can degrade quantum signals in coexistence scenarios. We evaluate out-of-band non-linear interference arising from \gls*{fwm} and \gls*{sprs}. For \gls*{sdm} systems, we account for spatial crosstalk and the appropriate scaling of non-linear effects within mode groups containing multiple degenerate modes. In counter-propagating configurations, we examine impairments due to Rayleigh backscattering and \gls*{sprs}. We also consider in-band distortions caused by \gls*{srs}. To unify these effects, we develop a semi-analytical model that integrates all the relevant physical impairments discussed. The model provides a set of differential equations that are able to account for interference arising from interactions of two or more physical effects, assuming the interference phenomena to be uncorrelated locally. Nevertheless, some approximations are provided to efficiently calculate the \gls*{fwm} contribution.

The paper is organized as follows. In Section~\ref{sec:nonlineareffects}, we discuss possible coexistence configurations and the most relevant interference effects that might be present. In Section~\ref{sec:model}, we develop our model to quantify the accumulated interference in a quantum channel with arbitrary center frequency, and introduce approximations for efficient numerical evaluation. In Section~\ref{sec:numericalevaluation}, we apply the model to sample scenarios, and discuss how different system parameters may affect the accumulated interference. Section~\ref{sec:conclusion} is devoted to conclusions.

% Section: 1
\section{Interference Effects Under Coexistence Schemes}
\label{sec:nonlineareffects}

\subsection{System Parameters for Coexistence and Impairment Impact}

Different interference effects become dominant under each coexistence scheme. The propagation direction relative to classical signals, frequency separation, and the use of \gls*{sdm} fibers to provide dedicated spatial channels for the quantum signal will change the magnitude and sources of interference contributions.

The quantum signal may be co-propagated with classical signals. However, this configuration requires steep roll-off filters to isolate the quantum channel at transmission. Alternatively, when the quantum signal is counter-propagated, filtering requirements at the quantum signal transmission are relaxed. Nevertheless, in this case, the quantum channel is susceptible to backscattered photons from the classical channels, which can induce significant interference at the quantum receiver, especially since the quantum signal is already attenuated by the link~\cite{patel2012coexistence}.

The spectral positioning of the quantum signal within the frequency grid significantly impacts non-linear interference levels and propagation losses. A large frequency separation from classical channels is desirable for minimizing interference, but operating outside the conventional C-band may require dedicated components and introduces higher attenuation, degrading the quantum signal integrity. On the other-hand, placing the quantum signal closer in frequency to the classical channels increases non-linear interference, which also impairs quantum information transmission rates.

Novel fiber designs with increased spatial dimensions are also under consideration for future optical networks, under the umbrella of \gls*{sdm}. Some spatial channels within \gls*{sdm} links exhibit strong coupling, as seen in the orthogonal polarizations of \glspl*{ssmf}. The strongly-coupled channels can be aggregated into distinct weakly-coupled groups of degenerate modes\footnote{The labelling of ``modes'' is not restricted to \glspl*{mmf}. The individual cores of \glspl*{mcf} can be weakly- or strongly-coupled, depending on the core spacing~\cite{saitoh2016multicore}. Under this consideration, we can then apply the same labelling and mathematical modeling from modes within~\glspl*{mmf} to cores within~\glspl*{mcf}.}. Currently, quantum signals must be confined to a mode group or core supporting only two orthogonally polarized fields\footnote{Some studies have been conducted on multi-mode quantum signal propagation~\cite{cozzolino2019high,amitonova2020quantum,beraza2024towards}, but the viability of such schemes remains under investigation.}. Higher-dimensional quantum signal transmission across weakly-coupled mode groups has been demonstrated using \glspl*{mcf}~\cite{bacco2022quantum,zahidy2024practical}. The small divergence in propagation speeds between cores allows these higher-dimensional states to propagate with minimal phase mismatch. However, accumulated interference within each weakly-coupled mode group can vary, requiring more detailed analysis of coexistence penalties, which falls outside the scope of this work.

In contrast, classical signals can be transmitted over mode groups with arbitrary mode counts using appropriate \gls*{mimo} equalization techniques. As shown in~\cite{antonelli2015modeling}, the generated non-linear interference in strongly-coupled modes scales with the inverse of the mode count and must be accurately accounted in a generalized model.

It is worth recalling that the analysis of performance degradation in the quantum channel caused by coexistence with classical transmission is only meaningful in unamplified fiber links. In fact, it is not possible to amplify quantum signals without corrupting the photon states, due to the non-cloning theorem, and quantum repeaters are distant from reaching deployment maturity~\cite{pirandola2017fundamental}. Currently, long-reach transmission of quantum signals is performed with trusted relay nodes~\cite{peev2009secoqc,stucki2011long,sasaki2011field}.

\subsection{Sources of Coexistence-Induced Interference}

\begin{figure*}[!t]
    \centering
    \includegraphics{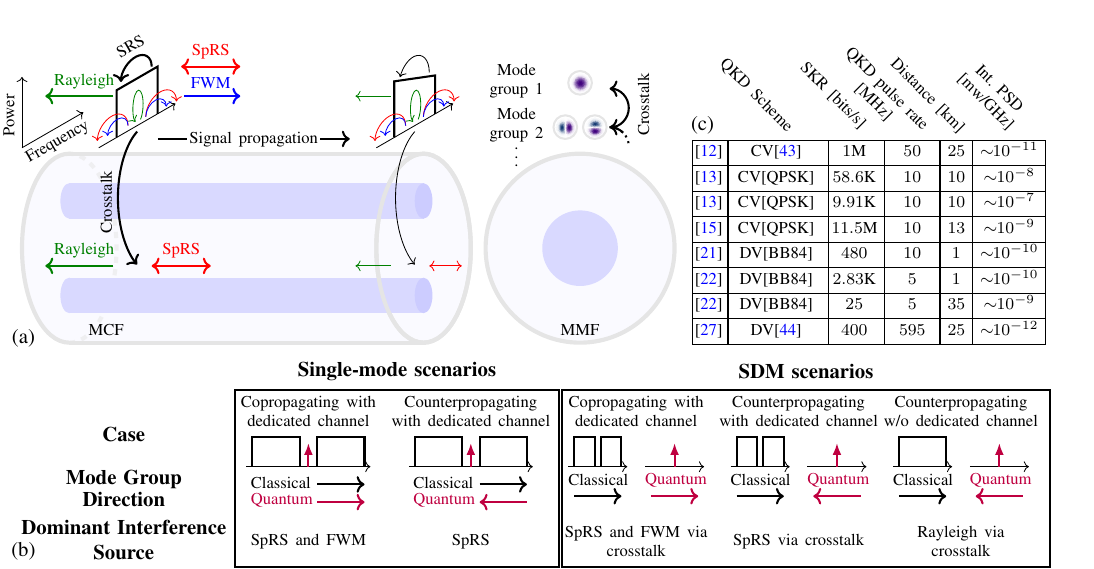}
    %\FigureDiagram
    \caption{(a) Power evolution of the classical signals and the predominant interference effects under coexistence schemes. The \gls*{fwm} generated from the crosstalked signal is disregarded due to its negligible intensity. \Glspl*{mcf} experience crosstalk from adjacent cores, and \glspl*{mmf} from distinct spatial mode groups. (b) Coexistence scenarios with corresponding signal directions and dominant interference effects on the quantum signal. In single-mode scenarios crosstalk is disregarded. (c) Reported \glspl*{skr} in various \gls*{qkd} implementations with interference \gls*{psd} values. For DV-\gls*{qkd}, explicit interference power values were not provided. Instead, the values were inferred from reported noise photon counts or from the \gls*{qber} and transmission rates, assuming the detector bandwidth equals the \gls*{qkd} symbol rate, unit detector efficiency, and a 50\% photon discard rate due to basis misalignment.\nocite{grosshans2003quantum,fung2006security}}
    \label{fig:Diagram}
\end{figure*}

In Fig.~\ref{fig:Diagram}(a), we present the main optical effects generated by classical signals that can contribute to interference at the quantum channel. \Gls*{sprs} results from the spontaneous decay of stimulated optical phonons in the silica lattice and can be viewed as a specific case of \gls*{srs}, where vacuum photons are amplified. The interference from \gls*{sprs} exhibits a broad bandwidth, extending up to~$\pm$40 THz from the pump frequency~\cite{lin2006raman}.

The effects of \gls*{srs} between classical signals can also be significant in high-power regimes, inducing power tilts in the classical signal profiles, which in turn may affect the interference contributions of the individual channels throughout propagation.

Interference from \gls*{fwm} can be significant when the quantum channel is closely spaced in frequency to the classical signals, generating out-of-band photons with power proportional to the cube of the classical signal powers~\cite{maeda1990effect}. Both \gls*{srs} and \gls*{fwm} generate photons along the direction of signal propagation, whereas \gls*{sprs} produces photons equally in both co- and counter-propagating directions~\cite{claps2002observation}.

Rayleigh scattering generates backward-propagating photons and can become a dominant source of interference in counter-propagating coexistence schemes. In addition, in \gls*{sdm} fibers with multiple weakly-coupled mode groups, spatial crosstalk must also be considered.

Figure~\ref{fig:Diagram}(b) illustrates the predominant sources of interference in selected coexistence schemes. Interactions among multiple effects can lead to higher-order interference. In \gls*{sdm} fibers, the quantum signal can be allocated such that direct interference becomes negligible, increasing isolation from classical channels, but  indirect interference via spatial crosstalk may still degrade the quantum signal. We neglect the effects of indirect \gls*{fwm}, as their magnitude scales exponentially faster than the interfering signal power. As a result, \gls*{fwm} generated by already weak interference noise signals is negligible compared to other indirect effects.

The key quantity to be evaluated in the analysis of coexistence between quantum and classical transmission is the interfering noise power. The noise power is directly related to common metrics utilized in \gls*{qkd} literature. In DV-\gls*{qkd} schemes, noise is typically considered in \gls*{qber} values. Under a single-photon encoding scheme and assuming homogeneous non-overlapping states, the \gls*{qber} can be expressed in terms of the noise photon count rate $N^{\Int}$ or interference noise power $P^{\Int}$, assuming unitary efficiency of the quantum detector~\cite[Eq.~(1)]{morrison2023single}
\begin{equation}
  \text{QBER}=\frac{N^{\Int}}{N^{\Sig}+N^{\Int}}=\frac{P^{\Int}}{N^{\Sig}hf+P^{\Int}},
  \label{eq:qber}
\end{equation}
where $N^{\Sig}$ is the received quantum signal rate, $h$ is Planck's constant, and $f$ is the optical frequency. In DV-\gls*{qkd}, all interference noise photons within the photon detector bandwidth contributes to erroneous measurements. In such schemes, $P^{\Int}$ is calculated as the integral of the noise \gls*{psd} over the detector bandwidth.

In CV-\gls*{qkd}, the secret fraction is determined by the mutual information Alice and Bob share minus the maximum information an eavesdropper (Eve) can gain on the key, impacted by the modulation variance, the reconciliation efficiency, the transmittance and the excess noise~\cite{laudenbach2018continuous}. The modulation variance can be optimized for any system with a variable optical attenuator, and the reconciliation efficiency depends on the chosen reconciliation scheme and the used error correcting code. The transmittance comprises the losses in the channel. Finally, the excess noise incorporates the practical impairments like detection noise, modulation noise, quantisation noise, and interference from the classical signals.

The excess noise ratio, expressed in \gls*{snu}, quantifies the additional noise variance beyond the vacuum noise, normalized to the shot noise level, given by ${\xi_{\text{excess}}=P^{\Int}/\sigma^{2}_{\eta\text{LO}}}$, where $\sigma^{2}_{\eta\text{LO}}$ is the local oscillator shot noise variance~\cite[Section~4]{laudenbach2018continuous}. The security impact of this excess noise is evaluated under the loose and strict security assumptions, commonly found in CV-\gls{qkd} literature. In the loose case, also known as the trusted-device model, we assume that all receiver related excess noise (detection noise and quantisation noise) is trusted and hence only affects the mutual information between Alice and Bob, but does not contribute to the information of Eve~\cite{usenko2016trusted, laudenbach2019analysis}. On the other hand, in the strict case, or untrusted-device model, we assume that Eve possesses the ability to characterize or exploit any imperfection in the receiver lab. Under this assumption, all excess noise is attributed to Eve, increasing her information and therefore lowering the secret fraction~\cite{laudenbach2018continuous}. The noises investigated in this paper originate from the channel, and are therefore always considered untrusted and fully attributed to Eve's information.

While several factors impact the \gls*{skr}, interference noise from classical channels is a primary limitation in coexistence scenarios. To define the tolerance limits for the coexistence interference, Fig.~\ref{fig:Diagram}(c) presents \gls*{skr} values from several previously reported \gls*{qkd} coexistence experiments, alongside the corresponding interference \gls*{psd} levels. For DV-\gls*{qkd}, the interference \gls*{psd} values are inferred using Eq.~\eqref{eq:qber}. The reported results indicate that CV-\gls*{qkd} systems can achieve non-zero \gls*{skr} with interference \gls*{psd} up to the scale of~${\sim\hspace{-3pt}10^{-7}}$~mW/GHz, while DV-\gls*{qkd} tolerates levels up to the scale of~${\sim\hspace{-3pt}10^{-9}}$~mW/GHz. The data suggest that coexistence is feasible up to these interference \gls*{psd} threshold values. As such, interference noise with \gls*{psd} values within this range should be explicitly incorporated into the model.

% Section: 2
\section{Power Equations of the Accumulated Interference}
\label{sec:model}

In this section, we derive power equations which quantify the accumulated interference within an arbitrary channel dedicated to a quantum signal. The derived formulas start from coupled-field equations for multi-mode propagation in \gls*{sdm} fibers supporting weakly-coupled mode groups of quasi-degenerate modes.

\subsection{Coupled-field equations}

Let $\Evec_{n,i}(z)$ denote the generalized Jones vector whose elements describe the complex envelopes of the degenerate modes belonging to mode group $n$ at frequency $f_{i}$. Establishing a reference frame that follows the intra-group random mode coupling, the evolution of the field vector obeys the following equation
\begin{equation}
  \frac{\partial\Evec_{n,i}(z)}{\partial z}=-\left[\frac{\alpha_{n,i}}{2}+j\beta_{n,i}\right]\Evec_{n,i}(z)+j\vec{\mathbf{N}}_{n,i}(z),
  \label{eq:nlse}
\end{equation}
where $\alpha_{n,i}$ and $\beta_{n,i}$ are, the mode-group- and frequency-dependent power attenuation coefficient and propagation constant, respectively, which (neglecting intra-group modal dispersion and \gls*{mdl}) are assumed to be identical for degenerate modes in the same mode group\footnote{In real scenarios there is a non-negligible deviation between degenerate modes attenuation, which results in \gls*{mdl}, and propagation constants, which results in inter-modal group dispersion. While some works have studied the statistics of such coefficients~\cite{ho2011statistics,antonelli2019stokes}, we assume such deviations to be negligible, and consider the coefficients as deterministic.}. The term $\vec{\mathbf{N}}_{n,i}(z)$ accounts for spatial-crosstalk between mode groups and Kerr non-linearities. In this subsection we neglect the effects of \gls*{srs}, which will be discussed in~\ref{ssec:srs}.

Using a perturbative approach, we can express the propagated field in the rotating reference frame as
\begin{equation}
  \Evec_{n,i}(z)=\Evec^{\Sig}_{n,i}(z)+\Evec^{\Int}_{n,i}(z),
  \label{eq:evecterms}
\end{equation}
where $\Evec^{\Sig}_{n,i}(z)$ is the unperturbed signal, and $\Evec^{\Int}_{n,i}(z)$ accounts for the accumulated additive interference. Since the interference field $\Evec^{\Int}_{n,i}(z)$ is the result of many stochastic interactions, it is uncorrelated with $\Evec^{\Sig}_{n,i}(z)$, and the total mode group power of each frequency channel can likewise be decomposed in
\begin{equation}
  \begin{split}
    P_{n,i}(z)&=\braket{||\Evec^{\Sig}_{n,i}(z)||^{2}}+\braket{||\Evec^{\Int}_{n,i}(z)||^{2}}\\
    &=P^{\Sig}_{n,i}(z)+P^{\Int}_{n,i}(z),
  \end{split}
\end{equation}
where $P_{n,i}(z)$, $P^{\Sig}_{n,i}(z)$, and $P^{\Int}_{n,i}(z)$ accounts for the total power over all degenerate modes in the $n^{\Th}$ mode group. In the remainder of this work, we assume that the mode group power is, on average, evenly spread across degenerate modes.

\subsection{Co-propagating interference contributions}

Under a co-propagating coexistence scheme, the main sources of interference are \gls*{sprs}, \gls*{fwm}, and spatial crosstalk. These effects result from uncorrelated physical processes and, within an infinitesimal fiber segment, their contributions can be evaluated independently. The local field perturbation $\vec{\mathbf{N}}_{n,i}(z)$ can then be defined as
\begin{equation}
  \begin{split}
    \vec{\mathbf{N}}_{n,i}(z)=&\overbrace{\sum_{m\neq n}\mathbf{K}^{(i)}_{nm}\Evec_{m,i}(z)g(z)}^{\text{Spatial crosstalk}}\hspace{-2pt}+\hspace{-2pt}\overbrace{\sum_{f_{h}\neq f_{i}}\boldsymbol{\zeta}^{(n)}_{ih}(z)\Evec_{n,h}(z)}^{\text{SpRS}}\hspace{-2pt}\\
    &\hspace{-2pt}+\underbrace{r_{n}\gamma_{n}\hspace{-12pt}\sum_{f_{h}-f_{k}+f_{l}=f_{i}}\hspace{-12pt}\left[\Evec_{n,h}(z)\cdot\Evec^{*}_{n,k}(z)\right]\Evec_{n,l}(z)}_{\text{FWM}},
  \end{split}
  \label{eq:nnli}
  \raisetag{10pt}
\end{equation}
where $\mathbf{K}^{(i)}_{nm}$ is the coupling matrix between degenerate modes of the $m^{\Th}$ and $n^{\Th}$ mode groups that characterizes the spatial crosstalk intensity, $g(z)$ accounts for perturbations in the waveguide boundary, $\boldsymbol{\zeta}^{(n)}_{ih}(z)$ is a random matrix, whose elements are uncorrelated, zero-mean, memoryless, complex random values, representing \gls*{sprs} from the $h^{\Th}$ frequency channel between degenerate modes of the $n^{\Th}$ mode group. The values of $\gamma_{n}$ and $r_{n}$ are, respectively for the $n^{\Th}$ mode group, the non-linearity coefficient\footnote{We define $\gamma_{n}$ as the fundamental mode non-linear coefficient, as widely employed in literature on discussions of non-linear interference~\cite{marcuse1997application,carena2012modeling}. The non-linearity coefficient is a function of the number of degenerate modes, and must be scaled accordingly~\cite{antonelli2015modeling}.} and a scaling factor, which is a function of the number of degenerate modes within the $n^{\Th}$ mode group.

The coupling matrix $\mathbf{K}^{(i)}_{nm}$ and the perturbation function are related to the mode-group-averaged spatial power coupling coefficient $\kappa^{(i)}_{nm}$ by~\cite[Eq.~(18)]{koshiba2011multi}
\begin{equation}
  \kappa^{(i)}_{nm}=\frac{1}{D_{m}}\Braket{\Tr\left[\left(\mathbf{K}^{(i)}_{nm}\right)^{H}\hspace{-4pt}\mathbf{K}^{(i)}_{nm}\right]}L_{C}S_{R}\left(L_{C}\Delta\beta^{(i)}_{nm}\right),
\end{equation}
where $D_{n}$ is the number of degenerate modes at the $n^{\Th}$ mode group, $S_{R}(\cdot)$ is the Fourier transform of the correlation function of the phase function describing waveguide deformations $g(z)$, with unitary correlation length, $L_{C}$ is the spatial crosstalk correlation length~\cite{koshiba2011multi}, and $\Delta\beta^{(i)}_{nm}=\beta_{n,i}-\beta_{m,i}$. The function $\Tr(\cdot)$ is the matrix trace, and the superscript $(\cdot)^{H}$ represents the Hermitian adjoint. The spectral function $S_{R}(\cdot)$ is known in closed-form for various correlation profiles in~\cite[Eq.~(19)-(21)]{koshiba2011multi}.

Even under weak coupling regimes, spatial crosstalk can result in non-negligible power depletion. Since the crosstalk-induced depletion is proportional solely to the power in the $(n,i)^{\Th}$ channel~\cite[Eq.~(29)]{marcuse1972derivation}, it is accounted into the attenuation coefficient $\alpha_{n,i}$ throughout this work.

The elements of the matrix $\boldsymbol{\zeta}^{(n)}_{ih}(z)$ are random in nature, with zero mean and mode-group-averaged variance equal to half the Raman cross-section captured by the waveguide in the $n^{\Th}$ mode group
\begin{equation}
  \frac{1}{D_{n}}\Braket{\Tr\left[\left(\boldsymbol{\zeta}^{(n)}_{ih}(z)\right)^{H}\hspace{-4pt}\boldsymbol{\zeta}^{(n)}_{ih}\right]}=\frac{\eta^{(n)}_{ih}}{2}.
\end{equation}

The Raman cross-section $\eta^{(n)}_{ih}$ can be expressed as a function of the Raman gain efficiency $g^{(n)}_{\R}(\Delta f)$~\cite{hellwarth1963theory},~\cite[Eq.~(6)]{bromage2004raman},~\cite[Eq.~(21)]{lin2007photon}
\begin{equation}
  \eta^{(n)}_{ih}=
  \begin{cases}
    \left(1+\Psi_{ih}\right)hf_{i}B_{s}g^{(n)}_{\R}(f_{h}-f_{i}),&f_{i}<f_{h},\\
    \Psi_{ih}hf_{i}B_{s}g^{(n)}_{\R}(f_{i}-f_{h}),&f_{i}>f_{h},
  \end{cases}
  \label{eq:etagr}
\end{equation}
where $h$ is Planck's constant, $B_{s}$ is the signal bandwidth, and $\Psi_{ih}$ is the phonon occupancy factor, given by~\cite[Eq.~(25.35)]{ashcroft1976solid}
\begin{equation}
  \Psi_{ih}=\left\{\exp\left[\frac{h\left|f_{i}-f_{h}\right|}{k_{B}T}\right]-1\right\}^{-1},
  \label{eq:phononoccupancy}
\end{equation}
where $k_{B}$ is Boltzmann's constant and $T$ is the waveguide temperature in Kelvin. The factor $\Psi_{ih}$ in~\eqref{eq:etagr} shows that \gls*{sprs} noise at high Stokes shift is proportional to the Raman gain efficiency, and vanishes at high anti-Stokes shifts. At zero Kelvin, $\Psi_{ih}$ vanishes, and the anti-Stokes component of the \gls*{sprs} disappears. Nevertheless, \gls*{sprs} at anti-Stokes frequencies is noticible at room temperature, in channels close to the pump. Equation~\eqref{eq:etagr} is in agreement with prior measurement of Raman cross-section values~\cite{mandelbaum2003raman,lin2007photon,eraerds2010quantum,kawahara2011effect,lin2019spontaneous,wu2025integration}.

\begin{figure*}[!t]
\normalsize
\setcounter{MYtempeqncnt}{\value{equation}}
\begin{equation}
  \begin{split}
    \frac{\der P^{\Sig}_{n,i}(z)}{\der z}=&-\underbrace{\alpha_{n,i}P^{\Sig}_{n,i}(z)},\\[-6pt]
    \frac{\der P^{\Int}_{n,i}(z)}{\der z}=&-\overbrace{\alpha_{n,i}P^{\Int}_{n,i}(z)}^{\text{Loss}}+\overbrace{\sum_{m\neq n}\kappa^{(i)}_{nm}P_{m,i}(z)}^{\text{Spatial crosstalk}}+\overbrace{\sum_{h\neq i}\eta^{(n)}_{ih}P_{n,h}(z)}^{\text{SpRS}}\\
        &+\frac{r^{2}_{n}\gamma_{n}^{2}}{D_{n}^{2}} \underbrace{\left[\sum_{\substack{h\neq i\\k=2h-i}}(\Phi_{n,h}+2)P^{2}_{n,h}(z)P_{n,k}(z)\rho^{(n)}_{ihkh}(z)\right. }_{\text{Degenerate FWM}}+\underbrace{\left. 2D_{n}\hspace{-4pt}\sum_{\substack{h\neq i,h\neq l\\k=h+l-i}}\hspace{-4pt}P_{n,h}(z)P_{n,k}(z)P_{n,l}(z)\rho^{(n)}_{ihkl}(z)\right]}_{\text{Non-degenerate FWM}}
  \end{split}
  \label{eq:peqs}
\end{equation}
\hrulefill
\vspace*{4pt}
\end{figure*}

The non-linearity coefficient scaling factor $r_{n}$ is given by~\cite[Eq.~(68)]{antonelli2015modeling}
\begin{equation}
  r_{n}\approx
  \begin{cases}
    1,&D_{n}=1,\\
    \frac{D_{n}}{D_{n}+1}\left[\frac{4}{3}(1-F_{\R})+\frac{3}{2}F_{\R}\right],&D_{n}>1,
  \end{cases}
\end{equation}
where $F_{\R}$ is the fraction of stimulated Raman contribution to the non-linear susceptibility coefficient, within a single-mode and single-polarization scenario, and it is approximately $0.18$ in conventional glass fibers~\cite{poletti2008description,poletti2009dynamics}. In the case of a single-mode mode group with polarization multiplexing, ${D_{n}=2}$ and the scaling factor is given by the classical Manakov coefficient for polarization multiplexed \glspl*{ssmf} transmission~(${r_{n}=(8+F_{\R})/9}$).

Substituting the local field perturbation $\vec{\mathbf{N}}_{n,i}(z)$ from~\eqref{eq:nnli} in~\eqref{eq:nlse}, we obtain the power evolution equations in~\eqref{eq:peqs}, as detailed in~\appref{app:a}. We obtain two distinct sets of equations, one describing the evolution of the unperturbed signal and the other the accumulating interference power. We assume evenly spaced channels, but a generalized solution for a flexible grid can be obtained by replacing all frequency indexes within the summations boundaries with the actual frequency values (${i\rightarrow f_{i}}$). In~\eqref{eq:peqs}, $\Phi_{n,i}$ is the excess kurthosis factor of the $(n,i)^{th}$ channel signal~\cite[Table I]{semrau2018gaussian}, and $\rho^{(n)}_{ihkl}(z)$ is the \gls*{fwm} efficiency factor, given by
\begin{equation}
  \rho^{(n)}_{ihkl}(z)=2\Real{\frac{1-e^{-\left(\frac{1}{2}\Delta\alpha^{(n)}_{ihkl}+j\Delta\beta^{(n)}_{ihkl}\right)z}}{\frac{1}{2}\Delta\alpha^{(n)}_{ihkl}+j\Delta\beta^{(n)}_{ihkl}}},
  \label{eq:rho}
\end{equation}
with
\begin{equation}
  \begin{split}
    \Delta\alpha^{(n)}_{ihkl}&=\alpha_{n,i}-\alpha_{n,h}-\alpha_{n,k}-\alpha_{n,l},\\
    \Delta\beta^{(n)}_{ihkl}&=\beta_{n,i}-\beta_{n,h}+\beta_{n,k}-\beta_{n,l}.
  \end{split}
  \label{eq:dalphadbeta}
\end{equation}

From the second-order Taylor series expansion of $\beta_{n,i}(\omega)$, $\Delta\beta^{(n)}_{ihkl}$ can be approximated by
\begin{equation}
  \Delta\beta^{(n)}_{ihkl}\approx 2\pi^{2}\beta^{(n)}_{2}\left(f^{2}_{i}-f^{2}_{h}+f^{2}_{k}-f^{2}_{l}\right),
\end{equation}
where $\beta^{(n)}_{2}$ is the group velocity dispersion parameter of the $n^{\Th}$ mode group.

While spatial crosstalk may induce noticeable power loss, depletion losses at the interfering signals due to \gls*{sprs} and \gls*{fwm} are negligible and disregarded in our modeling.

Solving equation~\eqref{eq:peqs} in a co-propagating scenario requires knowledge of all power levels at $z=0$. At the fiber input, the accumulated interference noise is $P^{\Int}_{n,i}(0)=0$, while the unperturbed signals are characterized by their launch power, given by $P^{\Sig}_{n,i}(0)=P^{\Tx}_{n,i}$.

Some fiber designs may experience non-negligible levels of inter-mode-group non-linear interference~\cite{kroushkov2013cross} due to overlap between mode fields. We disregard these contributions in the present analysis for simplicity, as they can be incorporated into additional \gls*{sprs} and \gls*{fwm} terms in $\vec{\mathbf{N}}_{n,i}(z)$, where the corresponding coefficients depend on the mode-group-averaged cross-effective area of the mode field profiles~\cite{antonelli2013raman,antonelli2015modeling}. We refer the reader to \appref{app:c} for a extension of the model to account for inter-mode-group non-linearities.

\subsection{Numerical solution optimization of FWM contributions}
\label{ssec:fwmoptimization}

The model given in~\eqref{eq:peqs} can be solved numerically. However, the \gls*{fwm} term entails fast-oscillating terms, with a period of~$2\pi/\Delta\beta^{(n)}_{ihkl}$, which can be on the order of centimeters. Numerical solutions of fast-oscillating derivatives can result in significant error accumulation, if step sizes are not sufficiently small to encompass the oscillations. This finer granularity can increase computational complexity beyond practical levels~\cite{iserles2004numerical,iserles2005numerical,deano2017computing}.

As the signal propagates, the \gls*{fwm} contribution quickly decays, and the oscillations becomes negligible. In addition, the summed contribution of many \gls*{fwm} terms tends to average out the oscillatory components, since each term exhibits a distinct oscillation period due to its unique value of $\Delta\beta^{(n)}_{ihkl}$. For these reasons, it is appropriate to replace the \gls*{fwm} contribution with an equivalent average representation that eliminates the oscillatory terms. This approximation allows the same numerical methods to be applied with coarser granularity, resulting in significantly reduced computation time. The steps followed in this section bear resemblance to Filon's method for the integration of oscilatory functions~\cite{filon1930iii, abramowitz1965handbook}.

In the remainder of this section, we focus on the \gls*{fwm} interference contribution. The propagation equation for the interfering field power, isolating the \gls*{fwm} terms, can therefore be expressed as
\begin{equation}
  \frac{\der P^{\Int}_{n,i}(z)}{\der z}=-\alpha_{n,i}P^{\Int}_{n,i}(z)+\frac{\der P^{\Int,\text{FWM}}_{n,i}(z)}{\der z},
\end{equation}
where $\der P^{\Int,\text{FWM}}_{n,i}(z)/\der z$ represents the \gls*{fwm} interference components given in~\eqref{eq:peqs}. The accumulated \gls*{fwm} interference power at position $z$ is then given by
\begin{equation}
  P^{\Int}_{n,i}(z)=e^{-\alpha_{n,i}z}\int_{0}^{z}e^{\alpha_{n,i}z'}\frac{\der P^{\Int,\text{FWM}}_{n,i}(z')}{\der z'}\der z'.
  \label{eq:pint}
\end{equation}

Since only the unperturbed signals generate significant levels of \gls*{fwm}, we assume the interfering power profiles as exponentially decaying (${P_{n,i}(z)\approx P^{\Tx}_{n,i}e^{-\alpha_{n,i}z}}$). Solving for $P^{\Int}_{n,i}(z)$ then yields
\begin{equation}
  \begin{split}
    P^{\Int}_{n,i}(z)\hspace{-2pt}=&\frac{4r^{2}_{n}\gamma^{2}_{n}}{D^{2}_{n}}\left[\sum_{\substack{h\neq i\\k=2h-i}}\hspace{-6pt}\frac{(\Phi_{n,h}+2)(P^{\Tx}_{n,h})^{2}P^{\Tx}_{n,k}\chi^{(n)}_{ihkh}(z)}{(\Delta\alpha^{(n)}_{ihkh})^{2}+4(\Delta\beta^{(n)}_{ihkh})^{2}}\right. \\
    &\left. \quad +2D_{n}\hspace{-8pt}\sum_{\substack{h\neq i,h\neq l\\k=h+l-i}}\hspace{-4pt}\frac{P^{\Tx}_{n,h}P^{\Tx}_{n,k}P^{\Tx}_{n,l}\chi^{(n)}_{ihkl}(z)}{(\Delta\alpha^{(n)}_{ihkl})^{2}+4(\Delta\beta^{(n)}_{ihkl})^{2}}\right]e^{-\alpha_{n,i}z},
  \end{split}
  \raisetag{16pt}
  \label{eq:fwmcomplete}
\end{equation}
where $\chi^{(n)}_{ihkl}(z)$ is proportional to the accumulated \gls*{fwm} interference without losses, and is given by
\begin{equation}
  \chi^{(n)}_{ihkl}(z)=e^{\Delta\alpha^{(n)}_{ihkl}z}-2e^{\frac{1}{2}\Delta\alpha^{(n)}_{ihkl}z}\cos\left(\Delta\beta^{(n)}_{ihkl}z\right)+1,
  \label{eq:chifwm}
\end{equation}
which relates to the \gls*{fwm} efficiency $\rho^{(n)}_{ihkl}(z)$ as
\begin{equation}
  \rho^{(n)}_{ihkl}(z)=\frac{4e^{-\Delta\alpha^{(n)}_{ihkl}z}}{(\Delta\alpha^{(n)}_{ihkl})^{2}+4(\Delta\beta^{(n)}_{ihkl})^{2}}\frac{\der\chi^{(n)}_{ihkl}(z)}{\der z}.
  \label{eq:rhochi}
\end{equation}

As seen in~\eqref{eq:chifwm}, the accumulated \gls*{fwm} oscillates with a period of $\Delta\beta^{(n)}_{ihkl}$. These oscillations are bounded from above and below by slowly varying envelopes. The upper and lower envelope are obtained by setting~${\Delta\beta^{(n)}_{ihkl}z=\pi}$ and~${\Delta\beta^{(n)}_{ihkl}z=0}$, respectively. The envelope boundaries of $\chi^{(n)}_{ihkl}(z)$ are given by
\begin{equation}
  \begin{split}
    \chi^{(n),\text{Max}}_{ihkl}(z)&=e^{\Delta\alpha^{(n)}_{ihkl}z}+2e^{\frac{1}{2}\Delta\alpha^{(n)}_{ihkl}z}+1,\\
    \chi^{(n),\text{Min}}_{ihkl}(z)&=e^{\Delta\alpha^{(n)}_{ihkl}z}-2e^{\frac{1}{2}\Delta\alpha^{(n)}_{ihkl}z}+1.
  \end{split}
  \label{eq:chilimits}
\end{equation}

The \gls*{fwm} envelope boundaries converge to each other as $z$ increases, indicating that the oscillatory \gls*{fwm} components become negligible over long distances. We can then use a linear average of the \gls*{fwm} boundaries to obtain the following approximation
\begin{equation}
  \tilde{\chi}^{(n)}_{ihkl}(z)\approx\frac{\chi^{(n),\text{Max}}_{ihkl}(z)+\chi^{(n),\text{Min}}_{ihkl}(z)}{2}=e^{\Delta\alpha^{(n)}_{ihkl}z}+1.
  \label{eq:chiavg}
\end{equation}

Substituting $\chi^{(n)}_{ihkl}(z)$ with~\eqref{eq:chiavg} in~\eqref{eq:rhochi}, we obtain the approximate \gls*{fwm} efficiency factor $\tilde{\rho}^{(n)}_{ihkl}$, without oscillatory terms
\begin{equation}
  \tilde{\rho}^{(n)}_{ihkl}\approx\frac{4\Delta\alpha^{(n)}_{ihkl}}{(\Delta\alpha^{(n)}_{ihkl})^{2}+4(\Delta\beta^{(n)}_{ihkl})^{2}},
  \label{eq:rhoavg}
\end{equation}
which is $z$-independent.

Note that using the approximated solution~\eqref{eq:chiavg} in~\eqref{eq:fwmcomplete} results in a non-zero value for the \gls*{fwm} interfering power at the fiber input
\begin{equation}
  \begin{split}
    \hspace{-2pt}\tilde{P}^{\Int}_{n,i}(0)\hspace{-2pt}=\hspace{-2pt}\frac{8r^{2}_{n}\gamma^{2}_{n}}{D_{n}^{2}}\hspace{-4pt}&\left[\sum_{\substack{h\neq i\\k=2h-i}}\hspace{-6pt}\frac{(\Phi_{n,h}+2)(P^{\Tx}_{n,h})^{2}P^{\Tx}_{n,k}}{(\Delta\alpha^{(n)}_{ihkh})^{2}+4(\Delta\beta^{(n)}_{ihkh})^{2}}\right. \\
    &\left. +2D_{n}\hspace{-10pt}\sum_{\substack{h\neq i,h\neq l\\k=h+l-i}}\hspace{-4pt}\frac{P^{\Tx}_{n,h}P^{\Tx}_{n,k}P^{\Tx}_{n,l}}{(\Delta\alpha^{(n)}_{ihkl})^{2}+4(\Delta\beta^{(n)}_{ihkl})^{2}}\right]\hspace{-4pt},
  \end{split}
\end{equation}
which should be accounted for in the initial conditions of numerical solutions to~\eqref{eq:peqs}, when using the approximated \gls*{fwm} efficiency factor in~\eqref{eq:rhoavg}.

\begin{table}[!t]
    \centering
    \begin{tabular}{|c|cc|c|}
        \hline
        \multirow{2}{*}{\textbf{Scenario}} & \multicolumn{1}{c|}{\makebox[1cm][c]{\textbf{Single-Mode}}} & & \multirow{5}{*}{\textbf{Unit}} \\\cline{2-3}
        & & & \\[-2.5mm]
        & \multicolumn{2}{c|}{\textbf{SDM}} & \\\cline{1-3}
        & & & \\[-2.5mm]
        \multirow{2}{*}{\textbf{Parameter}} & \multicolumn{2}{c|}{\makebox[1cm][c]{\textbf{Mode group (if SDM)}}} & \\\cline{2-3}
        & \multicolumn{1}{c|}{} & & \\[-2.5mm]
        & \multicolumn{1}{c|}{\makebox[1cm][c]{\textbf{Classical}}} & \makebox[1cm][c]{\textbf{Quantum}} & \\\hline
        & \multicolumn{1}{c|}{} & & \\[-2.5mm]
        Total launch power & \multicolumn{1}{c|}{25} & $-\infty$ & dBm \\
        Channel spacing $B_{s}$ & \multicolumn{2}{c|}{50} & GHz \\
        Channel count $N_{\text{ch}}$ & \multicolumn{2}{c|}{88} & \\
        Bandwidth & \multicolumn{2}{c|}{4.4 [1530-1565 nm]} & THz \\
        Temperature $T$ & \multicolumn{2}{c|}{300} & K \\
        \parbox[c]{3.2cm}{\centering Raman gain eff. peak $G_{\R}$} & \multicolumn{1}{c|}{0.4} & 0.35 & 1/W/km \\
        \parbox[c]{3.2cm}{\centering Raman gain eff. slope $c_{\R}$} & \multicolumn{1}{c|}{0.0286} & 0.025 & \hspace{-2pt}1/W/km/THz\hspace{-2pt} \\
        Degenerate modes $D$ & \multicolumn{2}{c|}{2 polarizations ($D\hspace{-2pt}=\hspace{-2pt}2$)} & \\
        Non-linear coefficient $\gamma$ & \multicolumn{2}{c|}{1.3} & 1/W/km\\
        Raman contribution factor $F_{\R}$ & \multicolumn{2}{c|}{0.18} & \\
        Non-linear scaling factor $r$ & \multicolumn{2}{c|}{0.91} & \\
        Group-velocity dispersion $|\beta_{2}|$ & \multicolumn{2}{c|}{21.7} & ps$^{2}$/km\\
        Excess kurthosis $\Phi$ & \multicolumn{1}{c|}{-1 [QPSK]} & 0 & \\
        \parbox[c]{3.2cm}{\centering Rayleigh~scattering~factor~$\Gamma$} & \multicolumn{2}{c|}{$10^{-4}$} & 1/km \\
        Span length $L_{s}$ & \multicolumn{2}{c|}{100} & km \\
        Simulation sections per span & \multicolumn{2}{c|}{100} & \\\hline
    \end{tabular}
    \caption{Simulation parameters.}
    \label{tab:parameters}
\end{table}

\begin{figure}[!t]
    \centering
    \includegraphics{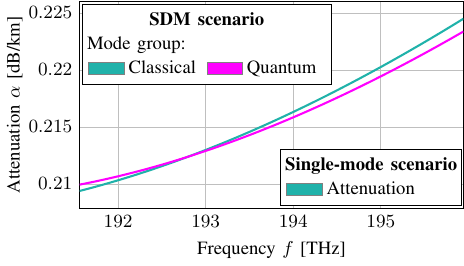}
    %\FigureAttenuation
    \caption{Frequency- and mode-group-dependent attenuation profiles for single-mode and \gls*{sdm} scenarios. The attenuation values are obtained using the model from~\cite{walker1986rapid}.}
    \label{fig:Attenuation}
\end{figure}

\begin{figure}[!t]
    \centering
    \includegraphics{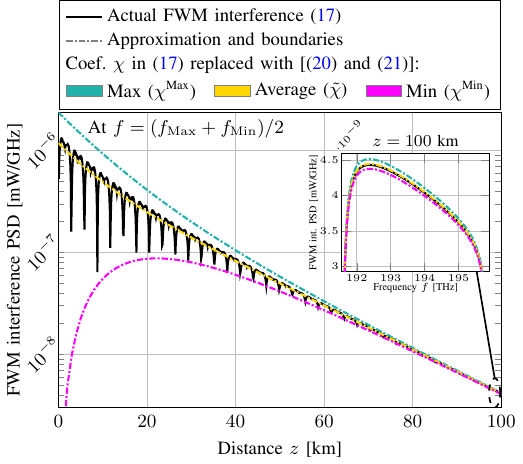}
    %\FigureFwmAverageApproximation
    \caption{\gls*{fwm} interference noise \gls*{psd} at the center frequency of the allocated spectrum (${f=(f_{\Max}+f_{\Min})/2}$) versus propagation distance. The solid line represents the analytical expression~\eqref{eq:fwmcomplete}, while the dot-dashed lines correspond to the solutions using the approximated expression from~\eqref{eq:chiavg} and the upper and lower envelope boundaries from~\eqref{eq:chilimits}. The inset graph shows the \gls*{fwm} interference noise \gls*{psd} at the fiber end. Simulation parameters are provided in Table~\ref{tab:parameters}, and attenuation values are shown in Fig.~\ref{fig:Attenuation} for a single-mode scenario.}
    \label{fig:FwmAverageApproximation}
\end{figure}

We validate the proposed approximation in a single-mode scenario with parameters given in Table~\ref{tab:parameters} and the attenuation profile shown in Fig.~\ref{fig:Attenuation}. The results are shown in Fig.~\ref{fig:FwmAverageApproximation}, where by solid line we plot the actual \gls*{fwm} interference noise \gls*{psd} over distance, given by~\eqref{eq:fwmcomplete}, and by dot-dashed lines we plot the solutions obtained by substituting $\chi^{(n)}_{ihkl}(z)$ with~\eqref{eq:chiavg}, as well as the corresponding upper and lower boundary envelopes from~\eqref{eq:chilimits}. The inset of Fig.~\ref{fig:FwmAverageApproximation} shows the \gls*{fwm} \gls*{psd} at the fiber end (${z=100}$~km). The solution using the proposed approximation is in close agreement to the actual \gls*{fwm} interfering power. The summed contribution of many \gls*{fwm} terms with distinct oscillation periods results in an averaging effect, which improves the agreement with the proposed approximation.

\subsection{FWM interference in the presence of SRS-induced tilt}
\label{ssec:srs}

Raman scattering in the form of \gls*{srs} is a common non-linear effect in wideband optical systems. \Gls*{srs} results in the transfer of power from higher- to lower-frequency channels. Although \gls*{srs} does not generate out-of-band non-linear interference, it distorts the power profiles during propagation, which in turn affects the levels of interference noise induced by other phenomena.

The power evolution of a signal, considering only losses and \gls*{srs}, is given by
\begin{equation}
  \frac{\der P_{n,i}(z)}{\der z}\hspace{-2pt}=\hspace{-2pt}-\alpha_{n,i}P_{n,i}(z)\hspace{-2pt}+\hspace{-2pt}\sum_{h\neq i}\hspace{-1pt}g^{(n)}_{\R}(f_{h}-f_{i})P_{n,i}(z)P_{n,h}(z),
  \label{eq:dpdzsrs}
\end{equation}
where we assume that all channels are sufficiently close in frequency that the energy conversion ratio can be approximated as unity (${f_{h}/f_{i} \approx 1}$).

As shown in~\cite{zischler2025closed}, the power profile of an arbitrary channel with \gls*{srs}-induced tilt can be approximated by
\begin{equation}
  P_{n,i}(z)\hspace{-2pt}=\hspace{-2pt}P_{n,i}(0)\exp\hspace{-2pt}\left[-\alpha_{n,i}z+c^{(n)}_{\R}(f^{(n)}_{\R}-f_{i})P_{\Total,n}L^{(n)}_{\text{eff}}(z)\right]\hspace{-2pt},
  \label{eq:psrs}
\end{equation}
where, for the $n^{\Th}$ mode group, $c^{(n)}_{\R}$ is the slope of the linear approximation of the Raman gain profile (${g^{(n)}_{\R}(f_{h}-f_{i})\approx (f_{h}-f_{i})c^{(n)}_{\R}}$), $f^{(n)}_{\R}$ is the zero-\gls*{srs}-induced tilt reference frequency, $P_{\Total,n}$ is the total launch power of the mode group, and $L^{(n)}_{\text{eff}}(z)$ is the total power profile effective length given by
\begin{equation}
  L^{(n)}_{\text{eff}}(z)=\frac{1-e^{-\alpha^{(n)}_{0}z}}{\alpha^{(n)}_{0}},
\end{equation}
where $\alpha^{(n)}_{0}$ is defined as the total power attenuation coefficient, where we approximate the total power as exponentially decaying (${\sum_{i=1}^{\Nch} P_{n,i}(z)\approx P_{\Total,n}e^{-\alpha^{(n)}_{0}z}}$). The values of $\alpha^{(n)}_{0}$ and $f^{(n)}_{\R}$ are found in~\cite{zischler2025closed} to be
\begin{equation}
  \begin{split}
    \alpha^{(n)}_{0}\hspace{-4pt}\approx& \sqrt[\mathrm{n}_{\R}\hspace{-1pt}]{\sum_{i=1}^{\Nch}\frac{\alpha^{\mathrm{n}_{\R}}_{n,i}P^{\Tx}_{n,i}}{P_{\Total,n}}},\\
    f^{(n)}_{\R}\hspace{-4pt}\approx&\frac{-1}{c^{(n)}_{\R}\hspace{-2pt} P_{\Total,n}L^{(n)}_{\text{eff}}\hspace{-2pt}(L_{s})}\hspace{-2pt}\ln\hspace{-4pt}\left[\sum_{i=1}^{\Nch}\hspace{-2pt}\frac{\alpha^{\mathrm{n}_{\R}}_{n,i}P^{\Tx}_{n,i}e^{[\alpha^{(n)}_{0}-\alpha^{(n)}_{n,i}]L_{s}}}{\left(\hspace{-2pt}\alpha^{(n)}_{0}\hspace{-2pt}\right)^{\hspace{-2pt}\mathrm{n}_{\R}}\hspace{-5pt}P_{\Total,n}e^{c^{(n)}_{\R}\hspace{-2pt}f_{i}P_{\Total,n}L^{(n)}_{\text{eff}}\hspace{-2pt}(L_{s})}}\hspace{-2pt}\right]\hspace{-4pt},
  \end{split}
  \raisetag{20pt}
\end{equation}
with $L_{s}$ denoting the span length, and $\mathrm{n}_{\R}$ being a non-zero positive integer fitting-parameter, whose best value can be shown to be 3 for C-band transmission~\cite{zischler2025closed}. An expression for~\eqref{eq:psrs} that accounts for inter-mode-group \gls*{srs} between weakly coupled mode groups is presented in~\cite[Eq.~(1)]{zischler2025evaluation} and further detailed in~\appref{app:c}.

From~\eqref{eq:psrs}, the combined effects of loss and \gls*{srs} can be described by a $z$-dependent effective loss coefficient given by
\begin{equation}
  \tilde{\alpha}_{n,i}(z)=\alpha_{n,i}-\frac{1}{z}c^{(n)}_{\R}(f^{(n)}_{\R}-f_{i})P_{\Total,n}(0)\Leff^{(n)}(z),
  \label{eq:alphatilde}
\end{equation}
where the unperturbed signals power profiles can then be described by
\begin{equation}
  P^{\Sig}_{n,i}(z)\approx P^{\Sig}_{n,i}(0)e^{-\tilde{\alpha}_{n,i}(z)z}.
\end{equation}

No analytical solution exists for the \gls*{fwm} efficiency factor when considering the \gls*{srs}-distorted power profiles in~\eqref{eq:psrs}, as further detailed in~\appref{app:b}. However, assuming that \gls*{srs} varies slowly with distance, such that the field evolution is primarily governed by the attenuation and propagation constants, the \gls*{fwm} efficiency factor can be approximated by
\begin{equation}
  \rho^{(n)}_{ihkl}(z)\approx2\Real{\frac{1-e^{-\left(\frac{1}{2}\Delta\tilde{\alpha}^{(n)}_{ihkl}(z)+j\Delta\beta^{(n)}_{ihkl}\right)z}}{\frac{1}{2}\Delta\tilde{\alpha}^{(n)}_{ihkl}(z)+j\Delta\beta^{(n)}_{ihkl}}},
  \label{eq:rhosrs}
\end{equation}
or, for the approximated solution given in~\eqref{eq:rhoavg}, by
\begin{equation}
  \tilde{\rho}^{(n)}_{ihkl}(z)\approx\frac{4\Delta\tilde{\alpha}^{(n)}_{ihkl}(z)}{(\Delta\tilde{\alpha}^{(n)}_{ihkl}(z))^{2}+4(\Delta\beta^{(n)}_{ihkl})^{2}},
  \label{eq:rhoappsrs}
\end{equation}
with
\begin{equation}
  \Delta\tilde{\alpha}^{(n)}_{ihkl}(z)=\tilde{\alpha}_{n,i}(z)-\tilde{\alpha}_{n,h}(z)-\tilde{\alpha}_{n,k}(z)-\tilde{\alpha}_{n,l}(z).
\end{equation}

\begin{figure}[!t]
    \centering
    \includegraphics{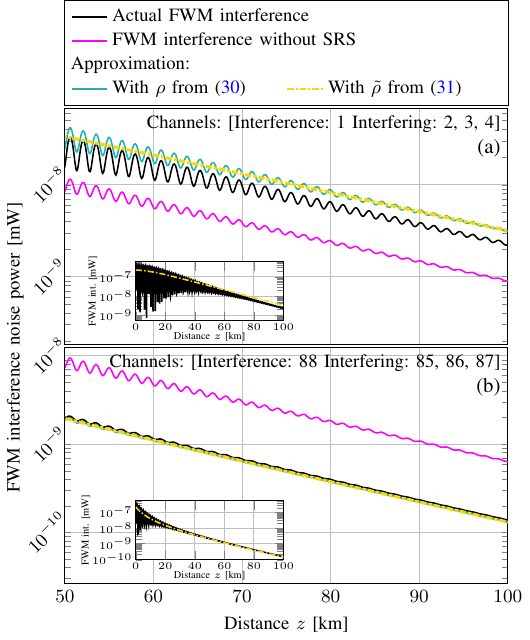}
    %\FigureFwmSrsApproximation
    \caption{Total \gls*{fwm} interference noise power in a 50~GHz channel at the (a) lower and (b) upper edges of the allocated spectrum versus distance in a single-mode scenario. The figure shows the actual \gls*{fwm} interference noise power, along with curves considering the proposed approximations for the \gls*{fwm} efficiency factor given in~\eqref{eq:rhosrs} and~\eqref{eq:rhoappsrs}. For reference, the \gls*{fwm} interference noise power neglecting \gls*{srs} is also presented. We consider only the strongest \gls*{fwm} contribution, arising from the three neighboring channels. A total launch power of 30~dBm is assumed, uniformly distributed across all channels. For the actual \gls*{fwm} curve, we utilize $10^{6}$ steps. The remaining simulation parameters are listed in Table~\ref{tab:parameters}, and the attenuation profile is shown in Fig.~\ref{fig:Attenuation}. To improve visual clarity, the curves are plotted for the second half of the fiber span. The inset illustrates the actual \gls*{fwm} interference and the approximation using~\eqref{eq:rhoappsrs} across the full span length.}
    \label{fig:FwmSrsApproximation}
\end{figure}

In Fig.~\ref{fig:FwmSrsApproximation}, we plot the \gls*{fwm} interference noise power with \gls*{srs} versus distance, along with the approximated solutions using~\eqref{eq:rhosrs} and~\eqref{eq:rhoappsrs}, for a single-mode scenario. We consider an extreme case of $30$~dBm total launch power evenly distributed across channels. The remaining parameters are listed in Table~\ref{tab:parameters}, and the attenuation profile is shown in Fig.~\ref{fig:Attenuation}. The actual values of the \gls*{fwm} efficiency factor $\rho^{(n)}_{ihkl}$ are obtained via numerical integration with the power profiles in~\eqref{eq:psrs}. We show only the \gls*{fwm} interference at the band edges, where approximation errors are expected to be most significant due to the increased magnitude of the \gls*{srs} term in~\eqref{eq:psrs}. The figure also includes the \gls*{fwm} interference in the absence of \gls*{srs}. The results illustrate that neglecting \gls*{srs} may result in significant discrepancies. On the other hand, the proposed approximations significantly improve estimation accuracy, while preserving high computational efficiency.

\subsection{Counter-propagating interference}
\label{ssec:back}

So far, we have discussed co-propagation interference under the assumption that all signals propagate along the positive $z$-axis. However, as previously noted, a counter-propagating quantum signal imposes less stringent isolation requirements. In such a configuration, the quantum signal is impaired by \gls*{sprs}, which is equally present in both propagation directions, and Rayleigh backscattering.

Assuming that all signals are allocated to distinct channels, we can trivially expand~\eqref{eq:peqs} to account for a set of counter-propagating signals by including an additional term for backscattered Rayleigh and \gls*{sprs}. As Rayleigh backscattering does not induce Stokes shifts, the scattered photons remain within the same frequency channel. The portion of Rayleigh backscattered light captured by the waveguide is characterized by the mode-group- and frequency-dependent Rayleigh scattering factor~$\Gamma_{n,i}$.

We can then generalize~\eqref{eq:peqs} to incorporate signal direction in the accumulating interference as
\begin{equation}
  \begin{split}
    \frac{d P^{\Fwd,\Int}_{n,i}(z)}{d z}=&\left. \frac{d P^{\Int}_{n,i}(z)}{d z}\right|_{\subalign{&P_{n,i}(z)=P^{\Fwd}_{n,i}(z)\\&P^{\Int}_{n,i}(z)=P^{\Fwd,\Int}_{n,i}(z)}}\hspace{-8pt}+\sum_{h\neq i}\eta^{(n)}_{ih}P^{\Bwd}_{n,h}(z)\\
    &+\Gamma_{n,i}P^{\Bwd}_{n,i}(z),\\
    -\frac{d P^{\Bwd,\Int}_{n,i}(z)}{d z}=&\left. \frac{d P^{\Int}_{n,i}(z)}{d z}\right|_{\subalign{&P_{n,i}(z)=P^{\Bwd}_{n,i}(z)\\&P^{\Int}_{n,i}(z)=P^{\Bwd,\Int}_{n,i}(z)}}\hspace{-8pt}+\sum_{h\neq i}\eta^{(n)}_{ih}P^{\Fwd}_{n,h}(z)\\
    &+\Gamma_{n,i}P^{\Fwd}_{n,i}(z),
  \end{split}
  \raisetag{10pt}
  \label{eq:peqsfull}
\end{equation}%
where the superscript $(\cdot)^{\Fwd}$ denotes signals propagating along the positive $z$-axis, and $(\cdot)^{\Bwd}$ denotes signals propagating in the opposite direction. The unperturbed signal power profiles, $P^{\Fwd,\Sig}_{n,i}(z)$ and $P^{\Bwd,\Sig}_{n,i}(z)$, can be obtained from the approximate closed-form solution given in~\eqref{eq:psrs}. The initial conditions are set at $z=0$ and $z=L_{s}$ for the forward- and backward-propagating signals, respectively. 

Solving~\eqref{eq:peqsfull} can be achieved with established numerical methods for multi-boundary ordinary differential equations~\cite{keller2018numerical}. However, if all classical signals propagate in the same direction, then the only counter-propagating contributions arise from Rayleigh and backscatter \gls*{sprs}, which have a negligible impact on the forward propagating interference. This simplification allows using more efficient numerical solvers, such as \gls*{rk4}, where first, we solve for the forward propagating interference, and with the resulting power profiles, we solve for the counter-propagating interference.

% Section: 3
\section{Results from Numerical Analysis}
\label{sec:numericalevaluation}

\begin{figure}[!t]
    \centering
    \includegraphics{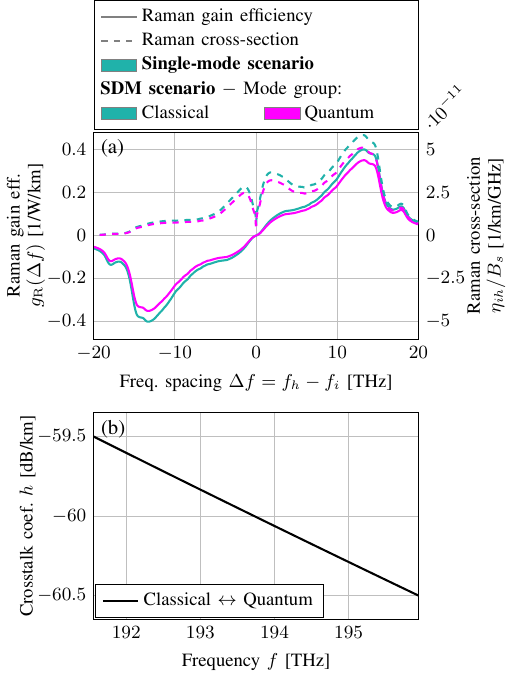}
    %\FigureProfiles
    \caption{Profiles of the coefficients for (a) Raman gain efficiency and spectral Raman cross-section, and (b) frequency-dependent crosstalk. The Raman gain profile is adapted from~\cite{stolen1989raman,lin2006raman}. The Raman cross-section is derived from~\eqref{eq:etagr} for a signal at the center frequency of the allocated spectrum (${f=(f_{\Max}+f_{\Min})/2}$), and is normalized by the signal bandwidth~$B_{s}$. The crosstalk is derived from~\cite[Eq.~(5)]{hayashi2011characterization}, for a linear slope of 1~dB/km across the allocated spectrum in wavelength scale, and crosstalk of $-60$~dB/km at the center wavelength.}
    \label{fig:Profiles}
\end{figure}

As discussed, distinct physical phenomena dominate interference generation under different scenarios. In this section, we employ numerical integration techniques to solve the proposed model and evaluate the interplay of the previously discussed interference effects for different coexistence setups. The profiles of the attenuation coefficient, the Raman gain efficiency and cross-section coefficients, and the crosstalk coefficient are shown in Fig.~\ref{fig:Attenuation}, Fig.~\ref{fig:Profiles}(a), and Fig.~\ref{fig:Profiles}(b), respectively. The frequency-independent parameters are listed in Table~\ref{tab:parameters}, and changes from the specified values are indicated alongside the corresponding results.

For the numerical evaluation, we select a single frequency slot where a quantum signal is to be allocated. For optimization purposes, we solve only for \gls*{sprs} and \gls*{fwm} interference affecting the frequency channel allocated to the quantum signal, and disregard \gls*{fwm} generation in the mode group of the quantum signal. For the unperturbed signals, we use the closed-form solution given in~\eqref{eq:psrs}. The interference is evaluated using Eqs.~\eqref{eq:peqs} and~\eqref{eq:peqsfull} for co-propagating and counter-propagating coexistence scenarios, respectively. The interference power evolution equations are solved using the \gls*{rk4} method, with the number of steps specified in Table~\ref{tab:parameters}. The numerical evaluation employs the approximation given in~\eqref{eq:rhoappsrs} for the \gls*{fwm} efficiency factor. In counter-propagating scenarios, we neglect the forward propagating interference generated by backscattered noise in order to utilize a single-boundary implementation of \gls*{rk4}\footnote{The implementation of the model used in this section is available online at~\url{https://gitlab.com/lucaszischler/quantum-coexistence-interference}.}

\subsection{Accuracy assessment}

To assess the accuracy of the proposed model, we consider a simplified single-mode scenario with 10 frequency channels allocated in the upper edge of the C-band and a total launch power of 10~dBm. The remaining parameters are listed in Table~\ref{tab:parameters}, while the attenuation and Raman profiles are shown in Fig.~\ref{fig:Attenuation} and Fig.~\ref{fig:Profiles}(a), respectively. The quantum signal is allocated in the highest frequency channel, and the launch power is evenly distributed among the classical channels. The numerically exact results are obtained by first computing the unperturbed signal power profiles under \gls*{srs} using the \gls*{rk4} method, solving the power evolution equations in~\eqref{eq:dpdzsrs} with the Raman gain profile from Fig.~\ref{fig:Profiles}(a). From the unperturbed signal power profiles, the $z$-dependent effective attenuation is computed as
\begin{equation}
  \tilde{\alpha}_{n,i}(z)=\frac{-1}{z}\ln\left[\frac{P^{\Sig}_{n,i}(z)}{P^{\Tx}_{n,i}}\right],
\end{equation}
which are then used to numerically integrate the \gls*{fwm} efficiency factors. The accumulated interference is obtained by solving the full power evolution model in~\eqref{eq:peqs}, again using \gls*{rk4}.

\begin{figure}[!t]
    \centering
    \includegraphics{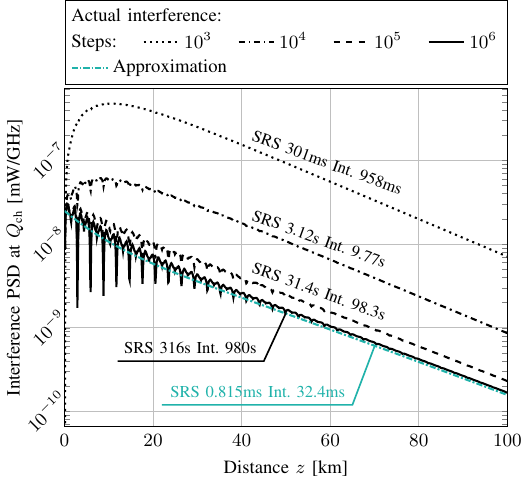}
    %\FigureAccuracy
    \caption{Interference curves for varying step counts and the proposed approximation in a single-mode scenario. The black curves show the numerically exact results with different step counts, indicated by line style, and the colored dot-dashed curve represents the proposed approximation using 100 steps. Elapsed computation times for the classical signal evolution under \gls*{srs} and interference evaluation are shown alongside each curve. The scenario considers 10 frequency channels in the upper C-band, with the quantum signal in the highest-frequency channel and 10~dBm of total launch power evenly spread among classical channels. Remaining parameters are listed in Table~\ref{tab:parameters}, while attenuation and Raman profiles are shown in Fig.~\ref{fig:Attenuation} and Fig.~\ref{fig:Profiles}(a). In the figure, $Q_{\text{ch}}$ denotes the quantum channel.}
    \label{fig:Accuracy}
\end{figure}

In Fig.~\ref{fig:Accuracy}, we plot the numerically exact accumulated interference computed using various step sizes against the proposed approximation. Execution times\footnote{Elapsed times correspond to a single run and are not representative of a rigorous benchmarking assessment.} required to evaluate the classical signal evolution under \gls*{srs} and to compute the corresponding interference values are displayed alongside the curves. The simulation was implemented in Python and executed on a Linux-based system equipped with a 12$^{\text{th}}$~Gen Intel i5-12450H processor. The results show that accurate evaluation of the full model requires a high number of integration steps to achieve convergence. Increasing the step count beyond $10^{6}$ results in minimal improvement in accuracy. In contrast, the proposed approximation achieves comparable accuracy with significantly fewer steps, reducing computational complexity while accurately matching the actual interference curve.

\subsection{Influence of launch power and channel spacing}

\begin{figure}[!t]
    \centering
    \includegraphics{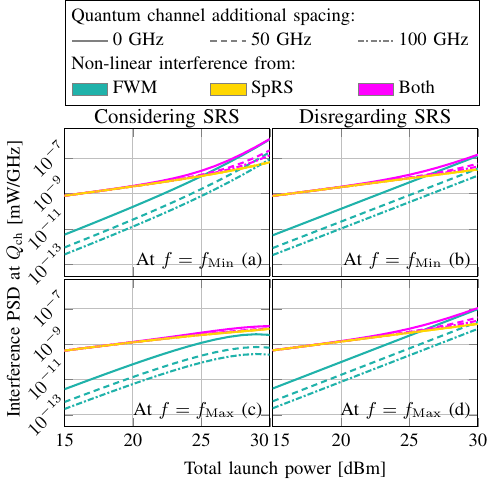}
    %\FigurePtxSweep
    \caption{Interference \gls*{psd} at the edges of the allocated spectrum ((a,~b) ${f_{\Min}}$ and (c,~d) ${f_{\Max}}$), evaluated at the fiber end (${z=100}$~km), as a function of total launch power. The curves in (a,~c) include \gls*{srs}, and in (b,~d) it is disregarded. Line styles represent different spacing values between quantum and classical signals.}
    \label{fig:PtxSweep}
\end{figure}

We start by evaluating the impact of launch power and quantum channel spacing, considering again a fully loaded C-band scenario. Figure~\ref{fig:PtxSweep} shows the accumulated interference versus total launch power at the fiber end, for a quantum channel placed at the edges of the spectrum in a single-mode scenario. In Figs.~\ref{fig:PtxSweep}(a,~c), we account for \gls*{srs}-induced distortion on the classical signals, while in Figs.~\ref{fig:PtxSweep}(b,~d) this effect is neglected. The different line styles indicate additional spacing between quantum and classical signals, implemented by deallocating classical channels adjacent to the quantum channel. The colors distinguish the interference contributions from \gls*{fwm}, \gls*{sprs}, and their combination. By comparing the results, we notice that \gls*{srs} influences interference values at launch powers greater than 25~dBm, especially in the \gls*{fwm} contribution.

In Fig.~\ref{fig:PtxSweep}, we also observe that \gls*{fwm} exhibits a stronger dependence on launch power than \gls*{sprs}, dominating the total interference at high power levels. The results for different channel spacing values indicate that \gls*{sprs} has a broader interference bandwidth, as its contribution remains nearly constant across the evaluated spacing values. In contrast, \gls*{fwm} decays sharply as spacing increases. This behavior is in line with the derived expression for the \gls*{fwm} efficiency factor, where the $\Delta\beta^{(n)}_{ihkl}$ term in the denominator of~\eqref{eq:rho} scales with the square of the frequency differences.

Additionally, we observe reduced interference at the highest-frequency channel compared to the lowest. For \gls*{sprs}, this is partly due to its weaker contribution at anti-Stokes frequencies, as illustrated in Fig.~\ref{fig:Profiles}(a). The interference also decays faster at higher frequencies as the channels near the upper edge of the spectrum experience greater attenuation. This effect is especially noticeable in the \gls*{fwm} contribution, which depends on the product of three interfering channel powers, as seen in~\eqref{eq:peqs}. \Gls*{srs} contributes to a further reduction in interference in the highest-frequency channels by transferring power from high to low frequency channels.

\subsection{Co-propagation in Single-Mode fibers}

\begin{figure}[!t]
    \centering
    \includegraphics{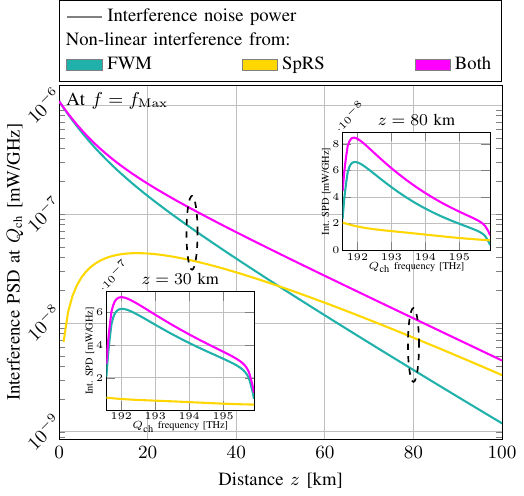}
    %\FigureTotalSmf
    \caption{Interference \gls*{psd} versus distance at the quantum channel, assuming the quantum signal is allocated at the upper edge of the spectrum. The insets show the interference \gls*{psd} versus the quantum channel frequency at 30~and 80~km. The colors shows the contributions from distinct non-linear interference sources.}
    \label{fig:TotalSmf}
\end{figure}

In this Subsection, we evaluate a single-mode scenario in which the quantum channel is allocated at a dedicated frequency slot, taking into account all previously discussed co-propagating physical phenomena, with the exception of spatial crosstalk. In Fig.~\ref{fig:TotalSmf}, we plot the accumulated interference \gls*{psd} at the quantum channel as a function of distance, considering the quantum signal placed at the highest-frequency channel. The different colors indicate the interference contributions from \gls*{fwm}, \gls*{sprs}, and their combination. In the insets of Fig.~\ref{fig:TotalSmf} we plot the interference \gls*{psd} versus the allocated quantum channel frequency at distances of 30~and 80~km.

From Fig.~\ref{fig:TotalSmf}, we notice that \gls*{fwm} results in significant interference near the fiber input but quickly decays owning to channel losses. In contrast, \gls*{sprs} interference accumulates gradually, reaching a peak before decaying as fiber losses begin to dominate the interference evolution, as previously reported in the literature~\cite{cai2020intercore}.

In the insets of Fig.~\ref{fig:TotalSmf}, we observe that at 30~km, \gls*{fwm} dominates the interference contribution across the entire spectrum. At 80~km, \gls*{sprs} becomes the dominant interference source at the highest-frequency channel, while \gls*{fwm} remains the main contributor elsewhere. The steep decline in \gls*{fwm} interference near the spectral edges is attributed to the absence of classical channels on one side of the quantum signal.

\subsection{Co-propagation in SDM fibers}
\label{ssec:copropagationsdmfiber}

As discussed, coexistence in \gls*{sdm} fibers can provide additional levels of isolation for the quantum signal. Here, we evaluate a scenario with two weakly-coupled mode groups, where one is allocated solely to the quantum signal and the other to classical transmission, accounting for all co-propagating physical phenomena. The frequency grid is the same for the quantum and classical mode groups. The quantum channel is allocated to a single frequency slot in the quantum mode group. The same frequency slot is left unused in the classical mode group. This is necessary to avoid in-band crosstalk, which would overwhelm the quantum signal at practical launch power and crosstalk levels.

\begin{figure}[!t]
    \centering
    \includegraphics{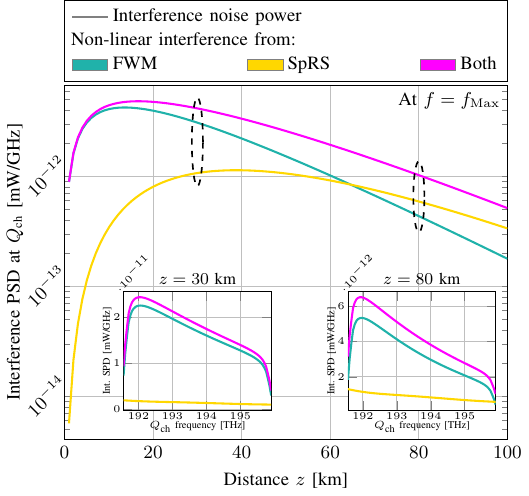}
    %\FigureTotalMmf
    \caption{Interference \gls*{psd} versus distance, considering the quantum channel to be allocated at the upper edge of the designated spectrum within a dedicated weakly-coupled mode group. The insets show the interference \gls*{psd} versus the quantum channel frequency at 30~and 80~km. The corresponding frequency is unallocated in the classical mode group to avoid in-band crosstalk. The colors shows the contributions from distinct non-linear interference sources.}
    \label{fig:TotalMmf}
\end{figure}

In Fig.~\ref{fig:TotalMmf}, we plot the accumulated interference \gls*{psd} as a function of distance. We notice a slower build-up near the fiber input, due to the isolation between mode groups. The insets of Fig.~\ref{fig:TotalMmf} show the interference \gls*{psd} versus the quantum channel frequency. Compared to the single-mode scenario in Fig.~\ref{fig:TotalSmf}, the additional mode group isolation reduces interference power values by approximately 40~dB.

\subsection{Counter-propagation in SDM fibers}

As discussed, this configuration avoids direct \gls*{fwm} interference. However, the quantum signal remains susceptible to Rayleigh and \gls*{sprs} backscattering, which may also mediate \gls*{fwm}. We evaluate a scenario similar to that in Subsection~\ref{ssec:copropagationsdmfiber}, now focusing on the \gls*{psd} of the backscattered interference, and accounting for all backscattering effects in this case. We consider the case where the quantum channel frequency is unallocated in the classical mode group. Additionally, we analyze a scenario in which a classical signal is present at the same frequency in the classical mode group.

\begin{figure}[!t]
    \centering
    \includegraphics{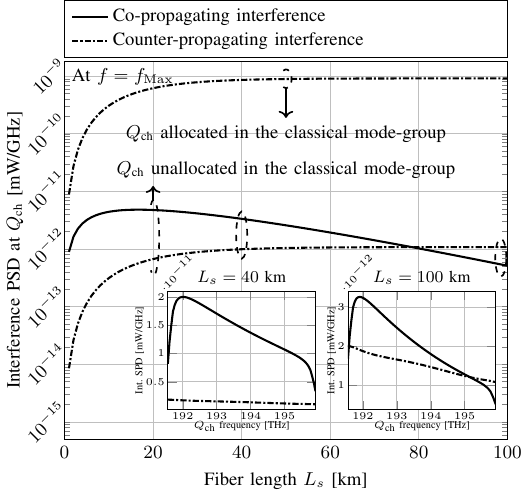}
    %\FigureTotalMmfBw
    \caption{Interference \gls*{psd} versus fiber length (${L_s}$) for co- and counter-propagating schemes, for a quantum channel allocated at the upper edge of the spectrum. Line styles indicate the propagation direction of the quantum signal. For counter-propagation, we compare scenarios with the quantum channel frequency allocated and unallocated in the classical mode group. Counter-propagating interference is evaluated at ${z=0}$, and co-propagating at ${z=L_{s}}$. The insets show interference \gls*{psd} versus quantum channel frequency at 40~km and 80~km, for the cases where the quantum channel is unallocated in the classical mode group.}
    \label{fig:TotalMmfBw}
\end{figure}

The counter-propagating interference profiles are calculated using the power derivatives given in~\eqref{eq:peqsfull}, neglecting their influence on the co-propagating signals, as discussed in Subsection~\ref{ssec:back}. The solutions are obtained via \gls*{rk4} and shown in Fig.~\ref{fig:TotalMmfBw}. In the figure, we plot the accumulated interference \gls*{psd} for a quantum signal at the highest-frequency channel versus the fiber length. The \gls*{psd} is evaluated at the fiber input ($z=0$) for counter-propagation and at the fiber output ($z=L_{s}$) for co-propagation. As expected, counter-propagating interference saturates after a certain fiber length. This occurs because most of the received backscattered light is generated near the classical signal transmitter. As the signals propagate, their power attenuates, resulting in reduced local backscattering. Additionally, the backscattered light must travel back to the quantum receiver, which further attenuates the contribution from regions far from the classical signal transmitter. Together, these effects explain the results showing that only backscattered light generated within a limited fiber length significantly contributes to the accumulated interference. Nevertheless, even if the noise contribution saturates, attenuation continues to increase exponentially with fiber length, reducing the power ratio between the received quantum signal and interference.

The insets in Fig.~\ref{fig:TotalMmfBw} show the interference \gls*{psd} versus the quantum channel frequency for fiber lengths of 40~km and 100~km, for the scenarios where the quantum channel frequency is unallocated in the classical mode group. At 40~km, co-propagating interference dominates over the entire spectrum, while at 100~km, counter-propagating interference becomes dominant at higher frequencies. Beyond the crossing point in Fig.~\ref{fig:TotalMmfBw} co-propagation results in lower interference than counter-propagation. For the plotted curves, this point occurs at fiber spans longer than 80~km. This crossing point is frequency-dependent, as seen in the insets. The optimal coexistence scheme will depend on the link length, along with factors such as frequency-dependent losses and device limitations.

The flatter spectral profile in the counter-propagating case reflects the absence of \gls*{fwm}, unlike in co-propagation, where the steep interference decay at the edges of the spectrum is an indication of the presence of \gls*{fwm}.

In the absence of a spectral notch at the quantum channel frequency in the classical mode group, interference increases by approximately 30~dB. When the notch is present, \gls*{sprs} dominates backscattered interference, either via crosstalk from classical mode group signals or through generation in the quantum mode group from crosstalked light. If a classical signal is present at the quantum frequency, backscattered interference is dominated by Rayleigh scattering into the quantum mode group via spatial crosstalk.

% Section: Conclusion
\section{Conclusion}
\label{sec:conclusion}

We presented a semi-analytical model to evaluate the accumulated interference impairing a quantum signal due to coexistence with classical transmission. The model accounts for the most significant physical phenomena, \gls*{sprs}, \gls*{fwm}, spatial crosstalk, and \gls*{srs}. While the solution is not closed-form, we provided numerical optimization techniques that allow for efficient and accurate evaluation.

Using the model, we analyzed selected coexistence scenarios. The results demonstrate that the uppermost frequency channel experiences the least interference, consistent with theoretical assumptions.

We showed that \gls*{fwm} can be substantial in co-propagating scenarios, especially in short-reach links, and its contribution can be orders of magnitude greater than that of \gls*{sprs}. Nevertheless, we observed that \gls*{fwm} is a narrowband effect, which can be mitigated by additional gigahertz-scale spacing between quantum and classical signals. In contrast, \gls*{sprs} is a wideband effect whose contribution does not significantly decrease with such narrow separation.

Depending on launch power and fiber length, different interference phenomena may dominate. Additionally, the choice between co- and counter-propagation of the quantum signal, relative to classical channels, influences the experienced interference levels.

Quantum channel placement is a multidimensional optimization problem. Nevertheless, the proposed model enables quick evaluation of a wide range of scenarios to optimize this choice.

% Appendices
\appendices
\def\sectionautorefname{appendix}

\section{Derivation of Power Evolution Equations}
\label{app:a}

The interference power evolution at the $(n,i)^{\Th}$ channel can be expressed as a function of the interference Jones field vector ${\Evec^{\Int}_{n,i}(z)}$ as
\begin{equation}
  \frac{\der P^{\Int}_{n,i}(z)}{\der z}\hspace{-3pt}=\hspace{-3pt}\Braket{\hspace{-2pt}\frac{\partial\left\Vert\Evec^{\Int}_{n,i}(z)\right\Vert^{2}}{\partial z}\hspace{-2pt}}\hspace{-3pt}=\hspace{-3pt}2\Real{\hspace{-2pt}\Braket{\hspace{-2pt}\frac{\partial\hspace{-2pt}\left(\hspace{-1pt}\Evec^{\Int}_{n,i}(z)\right)^{\hspace{-3pt}H}\hspace{-8pt}}{\partial z}\Evec^{\Int}_{n,i}(z)\hspace{-2pt}}\hspace{-2pt}}\hspace{-2pt},
  \label{eq:a1}
\end{equation}
where $(\cdot)^{H}$ is the conjugate transpose.

The unperturbed signal evolves according to
\begin{equation}
  \Evec^{\Sig}_{n,i}(z)=\Avec_{n,i}(0)e^{-(\frac{1}{2}\alpha_{n,i}+j\beta_{n,i})z},
\end{equation}
with $\Avec_{n,i}(z)$ denoting the complex wave amplitude vector. Equation~\eqref{eq:evecterms} implies the following decomposition
\begin{equation}
  \Avec_{n,i}(z)=\Avec_{n,i}(0)+\Avec^{\Int}_{n,i}(z),
\end{equation}
where $\Avec^{\Int}_{n,i}(z)$ is the complex wave amplitude accounting for the accumulated interference. We can then rewrite~\eqref{eq:a1} as a function of its respective complex wave amplitude ${\Avec^{\Int}_{n,i}(z)}$ as
\begin{equation}
  \frac{\der P^{\Int}_{n,i}(z)}{\der z}\hspace{-3pt}=\hspace{-3pt}-\alpha_{n,i}P^{\Int}_{n,i}\hspace{-1pt}(z)\hspace{-2pt}+\hspace{-2pt}2\mathbb{R}\hspace{-3pt}\left\{\hspace{-3pt}\Braket{\hspace{-2pt}\frac{\partial\hspace{-2pt}\left(\hspace{-1pt}\Avec^{\Int}_{n,i}(z)\hspace{-1pt}\right)^{\hspace{-3pt}H}\hspace{-8pt}}{\partial z}\Avec^{\Int}_{n,i}(z)\hspace{-2pt}}\hspace{-3pt}\right\}\hspace{-2pt}e^{\hspace{-2pt}-\alpha_{n\hspace{-1pt},\hspace{-1pt}i}z}\hspace{-2pt}.
  \label{eq:a2}
\end{equation}

The complex wave evolution is obtained from~\eqref{eq:nlse}, with~\eqref{eq:nnli}, and expressing the fields in terms of their complex wave components
\begin{equation}
  \begin{split}
    \frac{\partial\Avec^{\Int}_{n,i}\hspace{-1pt}(z)\hspace{-1pt}}{\partial z}\hspace{-3pt}=\hspace{-1pt}j\hspace{-3pt}\left\{\hspace{-2pt}\sum_{m\neq n}\hspace{-4pt}\mathbf{K}^{(i)}_{nm}\Avec_{m,i}(z)g(z)e^{\left(\frac{\alpha_{n\hspace{-1pt},\hspace{-1pt}i}-\alpha_{m\hspace{-1pt},\hspace{-1pt}i}}{2}+j(\beta_{n\hspace{-1pt},\hspace{-1pt}i}-\beta_{m\hspace{-1pt},\hspace{-1pt}i})\right)z}\right. &\\[-4pt]
    \hspace{8pt}+\hspace{-2pt}\sum_{h\neq i}\boldsymbol{\zeta}^{(n)}_{ih}(z)\Avec_{n,h}(z)e^{\left(\frac{\alpha_{n\hspace{-1pt},\hspace{-1pt}i}-\alpha_{n\hspace{-1pt},\hspace{-1pt}h}}{2}+j(\beta_{n\hspace{-1pt},\hspace{-1pt}i}-\beta_{n\hspace{-1pt},\hspace{-1pt}h})\right)z}&\\[-12pt]
    \hspace{8pt}+\hspace{-0.5pt}r_{n}\gamma_{n}\hspace{-12pt}\sum_{h-k+l=i}\left. \hspace{-12pt}\left[\hspace{-1pt}\Avec_{n,h}(z)\hspace{-2pt}\cdot\hspace{-2pt}\Avec^{*}_{n,k}(z)\hspace{-1pt}\right]\hspace{-3pt}\Avec_{n,l}(z)e^{\hspace{-2pt}\left(\hspace{-1pt}\frac{\Delta\alpha^{(n)}_{ihkl}}{2}\hspace{-1pt}+\hspace{-1pt}j\Delta\beta^{(n)}_{ihkl}\right)\hspace{-1pt}z}\hspace{-1pt}\right\}\hspace{-3pt},&
  \end{split}
  \label{eq:a3}
  \raisetag{-40pt}
\end{equation}
where $\Delta\alpha^{(n)}_{ihkl}$ and $\Delta\beta^{(n)}_{ihkl}$ are defined in~\eqref{eq:dalphadbeta}. The frequency matching condition (${h-k+l=i}$) assumes evenly spaced channels. Under a small-perturbation approximation, similar to~\cite{marcuse1972derivation}, we integrate~\eqref{eq:a3} with respect to $z$, from an arbitrarily close point $z_{0}$ (${z_0 \rightarrow z}$), resulting in
\begin{equation}
  \Avec^{\Int}_{n,i}(z)=\Avec^{\Int}_{n,i}(z_{0})+\int_{z_{0}}^{z}\frac{\partial\Avec^{\Int}_{n,i}(z')}{\partial z'}\der z',
  \label{eq:a4}
\end{equation}
where, under this small-perturbation assumption, we consider non-linear and crosstalk effects to remain uncorrelated over the interval ${(z_{0}, z)}$. We can then expand the expectation of the product between~\eqref{eq:a3} and~\eqref{eq:a4} given in~\eqref{eq:a2}\footnote{For mathematical manipulation, we rely on the following identities:
\begin{itemize}
  \item $\vec{\mathbf{x}}^{H}\mathbf{B}\vec{\mathbf{x}}=\Tr\left(\vec{\mathbf{x}}^{H}\mathbf{B}\vec{\mathbf{x}}\right)$.
  \item $\Tr\left(\mathbf{ABC}\right)=\Tr\left(\mathbf{BCA}\right)$.
  \item $\Braket{\mathbf{AB}}=\Braket{\mathbf{A}}\Braket{\mathbf{B}}$ if $\mathbf{A}$ and $\mathbf{B}$ are uncorrelated.
  \item $\Braket{\vec{\mathbf{x}}\vec{\mathbf{x}}^{H}}=\mathbf{I}$ if $\vec{\mathbf{x}}$ is an isotropic random vector. The Jones vector of degenerate modes $\Evec$ can be rewritten as $\Evec=\frac{1}{\sqrt{D}}\langle||\Evec||\rangle\vec{\mathbf{x}}$.
\end{itemize}}.

The term $\Braket{g(z')g^{*}(z)}$ in the expectation of the product of~\eqref{eq:a3} and~\eqref{eq:a4} gives the correlation of waveguide perturbations
\begin{equation}
  \Braket{g(z')g^{*}(z)}=R\left(\frac{z-z'}{L_{C}}\right),
\end{equation}
where $R(\cdot)$ is a correlation function shape with unit correlation length, and perturbation variance normalized to unity. Considering the correlation length $L_{C}$ of the mode-coupling perturbations sufficiently short, we can assume that the local spatial crosstalk contribution to the complex wave and $\Avec^{\Int}_{n,i}(z_{0})$ are uncorrelated. The solution for the spatial crosstalk terms can then be obtained from~\cite[Eq.~(18)]{koshiba2011multi}.

The \gls*{sprs} is a memoryless, zero-mean random process, therefore, its covariance can be expressed as an impulse response
\begin{equation}
  \frac{1}{D_{n}}\Braket{\Tr\left[(\boldsymbol{\zeta}^{(n)}_{ih}(z))^{H}\boldsymbol{\zeta}^{(n)}_{ih}(z')\right]}=\frac{\eta^{(n)}_{ih}}{2}\delta\left(z'-z\right),
\end{equation}
where, again, this term has no correlation with ${\Avec^{\Int}_{n,i}(z_0)}$, allowing its contribution to the power evolution to be derived independently.

The \gls*{fwm} effect is deterministic and depends on the propagated signals in other channels, making it inherently $z$-dependent. The complex wave $\Avec^{\Int}_{n,i}(z_0)$ can be decomposed into two components
\begin{equation}
  \Avec^{\Int}_{n,i}(z_{0})=\Avec^{\Int,\text{FWM}}_{n,i}(z_{0})+\Avec^{\Int,\text{Other}}_{n,i}(z_{0}),
\end{equation}
where $\Avec^{\Int,\text{FWM}}_{n,i}(z_{0})$ accounts for the first-order \gls*{fwm} generation in channel $(n,i)$, and $\Avec^{\Int,\text{Other}}_{n,i}(z_{0})$ results from \gls*{sprs}, crosstalk, and higher-order interactions. As \gls*{sprs} and spatial crosstalk are stochastic, these terms and their higher-order interactions are random and weakly correlated with the local \gls*{fwm} contribution at $z$. In contrast, $\Avec^{\Int,\text{FWM}}_{n,i}(z_{0})$ is correlated with the local \gls*{fwm} contribution and cannot be neglected. The total \gls*{fwm} contribution to the local interference power is given by

\begin{equation}
  \begin{split}
    \frac{\der P^{\Int,\text{FWM}}_{n,i}\hspace{-1pt}(z)}{\der z}\hspace{-3pt}=&2\mathbb{R}\hspace{-2pt}\left\{\hspace{-2pt}\left\langle\hspace{-0pt} \frac{\partial (\Avec^{\Int,\text{FWM}}_{n,i}(z))^{H}}{\partial z}\right. \right. \\
    &\left. \left. \hspace{16pt}\times\int_{0}^{z}\hspace{-3pt}\frac{\partial\Avec^{\Int,\text{FWM}}_{n,i}(z')\hspace{-3pt}}{\partial z'}\der z'\hspace{-0pt}\right\rangle\hspace{-2pt}\right\}\hspace{-1pt}e^{-\alpha_{n,i}z},
  \end{split}
  \label{eq:a6}
  \raisetag{12pt}
\end{equation}
where the \gls*{fwm} complex wave evolution term is given by
\begin{equation}
  \begin{split}
    \frac{\partial(\Avec^{\Int,\text{FWM}}_{n,i}(z))}{\partial z}=jr_{n}\gamma_{n}\hspace{-8pt}\sum_{h-k+l=i}\hspace{-8pt}\left[\hspace{-1pt}\Avec_{n,h}(z)\hspace{-2pt}\cdot\hspace{-2pt}\Avec^{*}_{n,k}(z)\hspace{-1pt}\right]\Avec_{n,l}(z)&\\
    \times e^{\hspace{-2pt}\left(\hspace{-1pt}\frac{\Delta\alpha^{(n)}_{ihkl}}{2}\hspace{-1pt}+\hspace{-1pt}j\Delta\beta^{(n)}_{ihkl}\right)\hspace{-1pt}z}\hspace{-2pt}.&
  \end{split}
  \label{eq:a7}
  \raisetag{12pt}
\end{equation}

To solve~\eqref{eq:a6}, we need to know the evolution of the interfering complex waves. We neglect any higher-order \gls*{fwm} generation arising from weak interference waves, as such interactions are negligible due to the cubic power dependence of the \gls*{fwm} process. Therefore, the dominant interfering fields in the \gls*{fwm} interaction are the classical signals, allowing the following approximation
\begin{equation}
  \Avec_{n,h}(z)\approx \Avec^{\Sig}_{n,h}(z)=\Avec_{n,h}(0).
\end{equation}

The dot product of the complex wave vectors in~\eqref{eq:a7} can be expanded into a sum over the degenerate mode components $A_{n,q,h}$, where ${q\in[1,D_{n}]}$. Replacing~\eqref{eq:a7} into~\eqref{eq:a6} and expanding, results in a summation over frequency components and degenerate modes
\begin{equation}
  \begin{split}
    \frac{\der P^{\Int,\text{FWM}}_{n,i}\hspace{-1pt}(z)}{\der z}\hspace{-3pt}=&r^{2}_{n}\gamma^{2}_{n}\hspace{-12pt}\sum_{\substack{h-k+l=i\\h'-k'+l'=i\\q,q'\in[1,D_{n}]}}\hspace{-12pt}2\mathbb{R}\hspace{-2pt}\left\{\left\langle A_{n,q,h}(0)A^{*}_{n,q,k}(0)\Avec_{n,l}(0)\right. \right. \\
    &\left. \qquad \times A^{*}_{n,q',h'}(0)A_{n,q',k'}(0)\Avec^{*}_{n,l'}(0)\right\rangle\\
    &\left. \qquad \times \int_{0}^{z}\hspace{-4pt}e^{\hspace{-2pt}\hspace{-1pt}\frac{\Delta\alpha^{(n)}_{ihkl}}{2}(z'+z)\hspace{-1pt}+\hspace{-1pt}j\Delta\beta^{(n)}_{ihkl}(z'-z)}\der z\right\}\hspace{-2pt}.
  \end{split}
  \label{eq:a9}
\end{equation}

Assuming the transmitted signals across distinct frequency and spatial channels to be uncorrelated, the expectation in~\eqref{eq:a6} results in non-zero values only under specific channel-matching conditions. Since the interfering signal at channel~$i$ has negligible power compared to the modulated classical signals, we ignore terms involving the frequency index~$i$. Within the same degenerate mode, non-zero contributions occur when ${h=h'=l=l'}$ and ${k=k'=2h-i}$, corresponding to the degenerate \gls*{fwm} case, and when ${h=h'}$, ${k=k'}$, ${l=l'}$, or when ${h=k}$, ${k'=h'}$, and ${l=l'}$, corresponding to the non-degenerate \gls*{fwm} solutions. In different degenerate mode in the same mode group, degenerate \gls*{fwm} solutions provide no additional contribution, but non-degenerate \gls*{fwm} terms exist for each ${q\in[1,D_{n}]}$ when ${q=q'}$. In fact, experimental results confirm the presence of non-degenerate \gls*{fwm} contributions from distinct modes~\cite{essiambre2013experimental}. There are no non-zero contributions when ${q\neq q'}$.

Assuming that the complex waves across all degenerate modes have identical statistical properties, the local \gls*{fwm} contribution to the interference noise power for any ${q\in [1,D_{n}]}$ is given by
\begin{equation}
  \begin{split}
    \frac{\der P^{\Int,\text{FWM}}_{n,i}\hspace{-1pt}(z)}{\der z}\hspace{-3pt}=\hspace{-2pt}r^{2}_{n}\gamma^{2}_{n}\hspace{-3pt}\left[\hspace{-2pt}\overbrace{\sum_{\substack{h\neq i\\k=2h-i}}\hspace{-8pt}\left\langle |A_{n,q,h}(0)|^{4}\right\rangle\hspace{-2pt} \left\langle |A_{n,q,k}(0)|^{2}\right\rangle\hspace{-2pt}\rho^{(n)}_{ihkh}}^{\text{Degenerate FWM}}\right. &\\
    \left. \quad \hspace{-8pt}+\overbrace{2D_{n}\hspace{-10pt}\sum_{\substack{h\neq i,h\neq l\\k=h+l-i}}\hspace{-8pt} \left\langle |A_{n,q,h}(0)|^{2}\right\rangle\hspace{-2pt}\left\langle |A_{n,q,k}(0)|^{2}\right\rangle\hspace{-2pt}\left\langle |A_{n,q,l}(0)|^{2}\right\rangle\hspace{-2pt}\rho^{(n)}_{ihkl}}^{\text{Non-degenerate FWM}}\right]\hspace{-3pt},&
  \end{split}
  \label{eq:a8}
  \raisetag{22pt}
\end{equation}
where the $2D_{n}$ factor in the non-degenerate component accounts for all non-degenerate \gls*{fwm} contributions. The integral over $z$ is replaced by the \gls*{fwm} efficiency factor~$\rho^{(n)}_{ihkl}$. Since the expectations in~\eqref{eq:a8} can be expressed in terms of power, we can rewrite~\eqref{eq:a3} using those quantities, as shown in~\eqref{eq:peqs}.

\section{Inter-mode-group non-linear interference}
\label{app:c}

In some fiber designs, such as \glspl*{mmf}, the mode fields overlap by a non-negligible amount, resulting in direct non-linear interference between distinct mode groups, even in the absence of spatial crosstalk. The inter-mode-group contributions of \gls*{sprs} and \gls*{fwm} must originate from frequency channels different from that of the quantum signal. Otherwise, the dominant impairment to the quantum signal is direct crosstalk. This supports the assertion that inter-mode-group non-linear interference is uncorrelated with spatial crosstalk at the quantum channel frequency. Consequently, these effects can be evaluated independently.

\subsection{Inter-mode-group SpRS}

Inter-mode-group \gls*{sprs} at the $n^{\Th}$ mode group is given by
\begin{equation}
  \frac{\der P^{\text{IMG-SpRS}}_{n,i}(z)}{\der z}=\sum_{\substack{m\neq n\\h\neq i}}\eta^{(nm)}_{ih}P_{m,h}(z),
  \label{eq:imgsprs}
\end{equation}
where $\eta^{(nm)}_{ih}$ represents the Raman cross-section captured by the $n^{\Th}$ mode group from light in the $m^{\Th}$ mode group. Since \gls*{sprs} can be regarded as a specific case of \gls*{srs}, both effects scale accordingly. The inter-mode-group Raman cross-section is given by ${\eta^{(nm)}_{ih} = \eta^{(n)}_{ih}\Aeff{,n}/\Aeff{,n,m}}$, where $\eta^{(n)}_{ih}$ is defined in~\eqref{eq:etagr}, $\Aeff{,n}$ is the $n^{\text{th}}$ mode-group-averaged effective area, and $\Aeff{,n,m}$ is the average cross-effective area between mode groups $n$ and $m$~\cite[Eq.~(14)]{antonelli2013raman},~\cite[Eq.~(2)]{zischler2025evaluation}.

\subsection{Inter-mode-group FWM}

Inter-mode-group \gls*{fwm} arises from the following component of the coupled Manakov equations~\cite[Eqs.~(2)~and~(3)]{mecozzi2012coupled}
\begin{equation}
  \frac{\partial \Evec^{\text{IMG-FWM}}_{n,i}(z)}{\partial z}=\gamma_{n}\hspace{-6pt}\sum_{\substack{m\neq n\\h-k+l=i}}\hspace{-6pt}r_{nm}\left[\Evec_{m,h}(z)\cdot \Evec^{*}_{m,k}(z)\right]\Evec_{n,l}(z),
\end{equation}
where the scaling factor $r_{nm}$ is proportional to the overlap integral between the degenerate modes of the $n^{\Th}$ and $m^{\Th}$ mode groups~\cite[Eq.~(4)]{mecozzi2012coupled},~\cite{antonelli2017nonlinear}.

The power evolution of the inter-mode-group \gls*{fwm} component can be derived in a similar approach to~\eqref{eq:a9}, with an additional summation index over the mode groups. We note that only non-degenerate \gls*{fwm} contributions are present in the inter-mode-group \gls*{fwm} term ($h\neq k\neq l\neq i$). The corresponding contribution to the interference power is described by the following expression
\begin{equation}
  \frac{\der P^{\text{IMG-FWM}}_{n,i}\hspace{-1pt}(z)}{\der z}\hspace{-2pt}=\hspace{-2pt}2\gamma^{2}_{n}\hspace{-11pt}\sum_{\substack{m\neq n\\h\neq i,h\neq l\\k=h+l-i}}\hspace{-12pt}\frac{r^{2}_{nm}}{D_{m}}\hspace{-1pt}P_{m,h}(z)P_{m,k}(z)P_{n,l}(z)\rho^{\hspace{-1pt}(nm)}_{ihkl}\hspace{-1pt}(z),
\end{equation}
where the inter-mode-group \gls*{fwm} efficiency factor $\rho^{(nm)}_{ihkl}(z)$ depends on the difference between attenuation and propagation constants of the distinct mode groups
\begin{equation}
  \begin{split}
    \Delta\alpha^{(nm)}_{ihkl}&=\alpha_{n,i}-\alpha_{m,h}-\alpha_{m,k}-\alpha_{n,l},\\
    \Delta\beta^{(nm)}_{ihkl}&=\beta_{n,i}-\beta_{m,h}+\beta_{m,k}-\beta_{n,l}.
  \end{split}
  \label{eq:danmdbnm}
\end{equation}

To reduce computational cost and incorporate the effects of \gls*{srs}, the efficiency factor can be approximated using~\eqref{eq:rhoappsrs}, with the corresponding coefficients given in~\eqref{eq:danmdbnm}.

\subsection{Inter-mode-group SRS}

Inter-mode-group \gls*{srs} results in increased spectral tilt due to power transfer between mode groups. As observed for \gls*{sprs}, the intensity of this contribution depends on the overlap between the field profiles of the different mode groups. The power profile of an arbitrary channel, accounting for inter-mode-group \gls*{srs} between weakly coupled groups of degenerate modes, can be approximated by~\cite[Eq.~(1)]{zischler2025evaluation}
\begin{equation}
  P_{n,i}(z)\hspace{-2pt}=\hspace{-2pt}P_{n,i}(0)e^{\hspace{-2pt}-\alpha_{n,i}z+\sum_{m} c^{(nm)}_{\R}(f^{(nm)}_{\R}-f_{i})P_{\Total,m}L^{(m)}_{\text{eff}}(z)\hspace{-2pt}},
  \label{eq:psrsimg}
\end{equation}
where $c^{(nm)}_{\R}$ is the inter-mode-group Raman gain efficiency slope, given by ${c^{(nm)}_{\R}=c^{(n)}_{\R}\Aeff{,n}/\Aeff{,n,m}}$, with ${c^{(nn)}_{\R}=c^{(n)}_{\R}}$. The value of $f^{(nm)}_{\R}$ can be approximated based on the energy conservation constraint of \gls*{srs}, which requires that ${P_{\Total,n}\approx\sum_{i=1}^{\Nch}P_{n,i}(z)e^{\alpha_{n,i}z}df}$.

The effective loss coefficient can be obtained from the exponent of~\eqref{eq:psrsimg}, in the same manner as in~\eqref{eq:alphatilde}, in order to utilize the approximation given in~\eqref{eq:rhoappsrs}.

\section{Derivation of FWM contribution with SRS}
\label{app:b}

From the derivative of~\eqref{eq:peqs} with respect to $z$, and under the assumption of an exponentially decaying total power, the power evolution considering only losses and \gls*{srs} is given by
\begin{equation}
  \frac{\der P_{n,i}(z)}{\der z}=\left[-\alpha_{n,i}+c_{\R}(f^{(n)}_{\R}-f_{i})P_{\Total,n}(z)\right]P_{n,i}(z).
  \label{eq:b1}
\end{equation}

As \gls*{fwm} does not affect the total power profile, and assuming it to be locally uncorrelated\footnote{While this is not strictly accurate, considering the Raman response of silica to be instantaneous within the time-scale of the transmitted pulses (The Raman response occurs within tenths of a picosecond~\cite{lin2006raman}. For conventional gigabaud transmission, the time dependence of the Raman response can be disregarded), the Raman contribution to the \gls*{fwm} term is already accounted within the factor $F_{\R}$~\cite{antonelli2015modeling},~\cite[Chapter~2]{agrawal2007nonlinear}.} with \gls*{srs}, the combined effect of both phenomena is expressed as
\begin{equation}
  \begin{split}
    \frac{\der P^{\Int}_{n,i}(z)}{\der z}=&\left[-\alpha_{n,i}+c_{\R}(f^{(n)}_{\R}-f_{i})P_{\Total,n}(0)e^{-\alpha_{0}z}\right]P^{\Int}_{n,i}(z)\\
    &+\frac{\der P^{\Int,\text{FWM}}_{n,i}(z)}{\der z},
  \end{split}
  \label{eq:b2}
\end{equation}
where the \gls*{fwm} contribution is defined in~\eqref{eq:a6}.

Equation~\eqref{eq:b2} is a first-order ordinary differential equation and can be solved as
\begin{equation}
  P^{\Int}_{n,i}(z)=e^{-\tilde{\alpha}_{n,i}(z)z}\int_{0}^{z}e^{\tilde{\alpha}_{n,i}(z')z'}\frac{\der P^{\Int,\text{FWM}}_{n,i}(z')}{\der z'}\der z',
  \label{eq:b3}
\end{equation}
where $\tilde{\alpha}_{n,i}(z)$ is given by~\eqref{eq:alphatilde}. Since \gls*{srs} is a phase-matched effect—due to the strong coupling between vibrational and Stokes waves~\cite{hellwarth1963theory}, it results in coherent wave accumulation, affecting only the field amplitude. The wave field can then be expressed as
\begin{equation}
  \Evec_{n,i}(z)=\Avec_{n,i}(z)e^{-\left\{\frac{1}{2}\left[\alpha_{n,i}-c_{\R}(f^{(n)}_{\R}-f_{i})P_{\Total,n}(z)\right]+\beta_{n,i}\right\}z},
  \label{eq:b4}
\end{equation}
assuming that all degenerate modes experience the same level of \gls*{srs}~\cite{antonelli2013raman}.

An accurate solution to~\eqref{eq:b3} requires the use of the field profiles in~\eqref{eq:b4}, which leads to a double integration when substituting the \gls*{fwm} term from~\eqref{eq:a6} into~\eqref{eq:b3}.

% Acknowledgment
%\section*{Acknowledgment}
% No acknowledgment

% Bibliography
\bibliographystyle{IEEEtran}
\bibliography{references}

\end{document}